\documentclass[prd]{revtex4}
\usepackage{epsfig}
\usepackage{mathrsfs}
\usepackage{amsmath}
\usepackage{amssymb}
\usepackage{color}
\usepackage{graphicx}

\begin{document}

\title{Resolved and unresolved Galactic gamma--ray sources}

\author{Paolo Lipari}

\email{paolo.lipari@roma1.infn.it}
\affiliation{INFN, Sezione Roma ``Sapienza'',
 piazzale A.Moro 2, 00185 Roma, Italy}

\affiliation{
  IHEP, Key Laboratory of Particle Astrophysics,
  Chinese  Academy of Sciences, Beijing, China}

\author{Silvia Vernetto}
\email{vernetto@to.infn.it}
\affiliation{
INAF, Osservatorio Astrofisico di Torino,
 via P.Giuria 1, 10125 Torino, Italy}

\affiliation{
 INFN, Sezione Torino,
 via P.Giuria 1, 10125 Torino, Italy}

\begin{abstract}
The Galactic gamma--ray flux can be described as the sum of two components:
the first is due to the emission from an ensemble of discrete sources,
and the second is formed by the photons produced 
by cosmic rays propagating in interstellar space and interacting with
gas or radiation fields.
The source component is partially resolved
as the contributions from individual sources,
but a fraction is unresolved and appears as
a diffuse flux. Both the unresolved source flux
and the interstellar emission flux 
encode information of great significance for high energy astrophysics,
and therefore the separation of these two contributions
is very important. 
In this work we use the distributions in celestial coordinates
of the objects contained in the catalogs obtained
by the Extensive Air Showers telescopes HAWC and LHAASO
to estimate the total luminosity of the Galactic gamma--ray sources
and the contribution of unresolved sources to the diffuse gamma--ray flux.
This analysis suggests that while the flux from unresolved sources
is measurable and important, 
the dominant contribution to the diffuse flux
over most of the celestial sphere is interstellar emission.
\end{abstract}

\maketitle

%\clearpage

\section{Introduction}
\label{sec:introduction} 

The study of the fluxes of high energy gamma--rays arriving at
Earth
from different astrophysical sources is of fundamental importance
for multi--messenger astrophysics, and is crucial for advancing our
understanding of the ``High Energy Universe''.
Recent years have seen a remarkable progress in this field,
with observations now covering an energy range of
more than seven orders of magnitude from less than $100$~MeV to more than
1~PeV ($10^8$--$10^{15}$~eV).
The observed gamma--ray flux  is composed of Galactic and extragalactic components,
but for $E \gtrsim 1$~TeV the extragalactic flux 
is strongly suppressed by absorption during propagation
and becomes negligible.
In this paper we will study this high energy range and
discuss only the Galactic component.

The observed Galactic gamma--ray flux is formed
by a component due to an ensemble of well localized sources
(point--like or quasi point--like) and a diffuse component 
smoothly distributed over the sky.
This diffuse component is due in part to the emissions from cosmic rays
propagating in interstellar space and
interacting with gas and radiation fields,
and in part to the sum of the contributions from those
sources that are too faint to be identified by the observations.

Decomposing the diffuse flux into these two components
(interstellar emission and unresolved sources)
and measuring their (different) spectral shapes
and angular distributions
is a very important task, because both components encode
very valuable information.
The measurement of the interstellar emission allows
to study the shape and the normalization of the
cosmic ray spectra in different regions of the Milky Way,
and on the other hand, a complete study of Galactic gamma--ray sources
must include a good understanding also of the faint, unresolved sources.

It should be noted, however, that the separation of the gamma--ray emission
into two components could become ambiguous in some cases, especially at high energy.
This is because it is possible, and indeed expected,
that cosmic ray particles accelerated in compact sources
will escape and form  halos of different sizes around
their accelerator. Gamma--rays are produced by the interactions of these
relativistic particles both inside the accelerators and in the halos,
so that a clear separation between the source and the interstellar medium
may be difficult.

In the GeV energy range the gamma--ray flux generated by interstellar emission
is significantly larger than the flux of unresolved sources,
however the two components have different spectral shapes, and
therefore the situation could become different at higher energy,
in the multi--TeV and PeV energy ranges.

The gamma--rays sky has now been observed by several telescopes,
and these observations have resulted in several source catalogs.
In the present work we study these catalogs in order to estimate
the properties of the populations of Galactic gamma--ray sources,
and then to evaluate the unresolved source flux.
Our approach here will be purely
phenomenological, in fact essentially geometrical,
and will not rely on the construction of a model for the nature of the
sources and for their emission mechanisms. Such a modeling
is of course of great interest, but is also very difficult,
since it is likely that different classes of sources contribute
to the gamma--ray flux.

This paper is organized as follows.
In the next section we describe the catalogs of high energy
($E \gtrsim 1$~TeV) gamma--ray sources obtained by the
Extensive Air Shower (EAS) telescopes HAWC and LHAASO,
and discuss the conditions for resolving a source.
The following section summarizes the existing
high energy measurements
of the diffuse gamma--ray flux.
Section~\ref{sec:method} introduces the geometrical method
we use in this paper to  derive the main properties of the
Galactic gamma--ray sources from a study of the sky distribution of
the resolved sources.
Section~\ref{sec:data_analysis} applies the geometrical method
to the HAWC and LHAASO catalogs 
obtaining estimates of the luminosity of
the individual sources and of the ensemble of all Galactic sources.
These results are then used to calculate the flux of unresolved
sources in the sky regions where the diffuse flux has been measured.
Section~\ref{sec:neutrinos} briefly discusses the  results on Galactic neutrinos
recently obtained by IceCube and the relationship between the gamma--ray
and neutrino fluxes.
The last section contains a summary and a final discussion.

\section{Gamma--ray source catalogs}
\label{sec:catalogs}
Catalogs of gamma--ray sources at high energies ($E \gtrsim 1$~TeV)
have been obtained by the Cherenkov telescopes HESS \cite{HESS:2018pbp} 
and VERITAS \cite{Acharyya:2023llt}, but for reasons that
will be described below, this work is based on the
catalogs obtained by two Extensive Air Shower (EAS) telescopes:
the High Altitude Water Cherenkov Observatory (HAWC),
located in Mexico at latitude $\lambda = 19.0^\circ$N
and altitude 4100~m,
and the Large High Altitude Air Shower Observatory (LHAASO),
located in China at latitude $\lambda = 29.36^\circ$N and altitude 4400~m.
\begin{itemize}
\item The third HAWC Catalog (3HAWC) \cite{HAWC:2020hrt}
lists 65 sources, observed in the 0.1--100~TeV energy range during 1523 days
of data acquisition (between November 2014 and June 2019).

\item The first LHAASO catalog (1LHAASO) \cite{LHAASO:2023rpg}
lists a total of 90 sources.
The telescope consists of two arrays: the water Cherenkov detector array (WCDA),
sensitive in the energy range (1--30)~TeV,
and the Kilometer Squared Array (KM2A)
sensitive in the range from approximately 25~TeV to more than 1~PeV.
Of the 90 sources in the catalog,
69 were observed by WCDA (in 508 days of data taking),
and 75 by KM2A (in 933 days of data taking),
with 54 sources identified in both instruments.
Out of the 75 sources observed by KM2A, 44 have measurable fluxes also for $E > 100$~TeV.
\end{itemize}

The present work discusses Galactic sources, and therefore the few
known extragalactic object contained in the catalogs have been excluded.
From the HAWC catalog we have excluded the two sources 
associated with the Markarian 421 and Markarian 501 blazars.
From the LHAASO catalog we have excluded five
extragalactic sources (all observed only in the WCDA array).
Four of them are associated with known extragalactic objects 
(the two Markarians and the blazars 1ES 2344+514 and 1ES 1727+502),
in addition also the source LHAASO~J1219+2915 was
classified as extragalactic in \cite{LHAASO:2023rpg}.
The number of additional, non--identified, extragalactic
objects in the two catalogs is expected to be very small.

\begin{figure}
\begin{center}
\includegraphics[width=14cm]{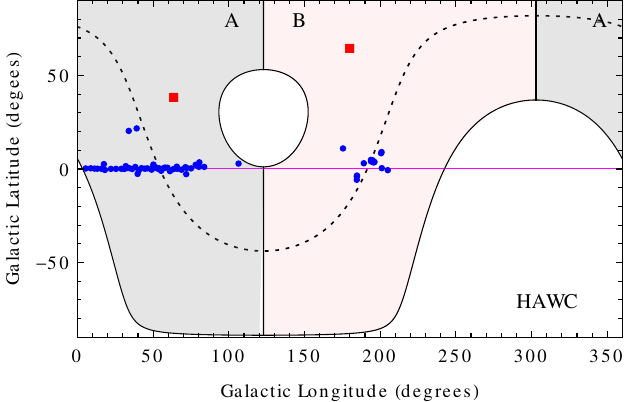}

\vspace{1.5cm}
\includegraphics[width=14cm]{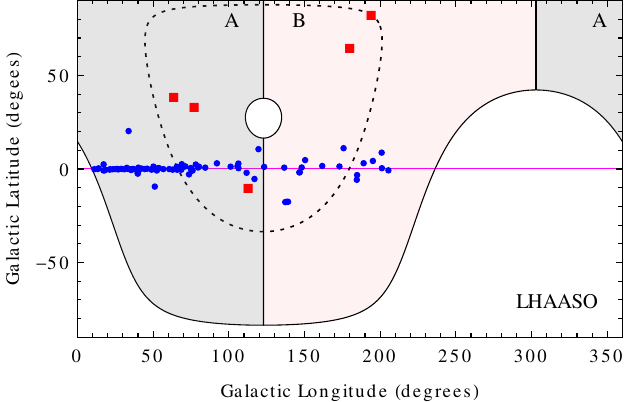}
\end{center}

\caption {\footnotesize
The two panels show the position of the gamma--ray sources in the
HAWC (top) \cite{HAWC:2020hrt} and LHAASO (bottom) \cite{LHAASO:2023rpg}
catalogs in Galactic coordinates.
The red squares are sources identified as extragalactic.
The shaded areas indicate the regions of the celestial sphere
that are observable given the positions of the two telescopes
(geographic latitude $\lambda = 19.0^\circ$ for HAWC and
$29.36^\circ$ for LHAASO) and assuming that the maximum angle
for observing the sources is $\theta_{\rm max} = 45^\circ$ for HAWC and
50$^\circ$ for LHAASO.
The different shadings identify the two subregions~A and~B [see Eq.~(\ref{eq:acc2})]
where the sensitivity for gamma--ray observations are equal.
The dashed line indicates the  points of celestial declination
equal to the telescope geographic latitude ($\delta = \lambda$)
that corresponds to the points with trajectories that transit through the local zenith.
%The vertical lines correspond to celestial right ascension $\alpha_1 =-167.14^\circ$
%and $\alpha_2 = +12.86^\circ$.
\label{fig:lhaaso_sky}}
\end{figure}

\begin{figure}
\begin{center}
\includegraphics[height=4.3cm]{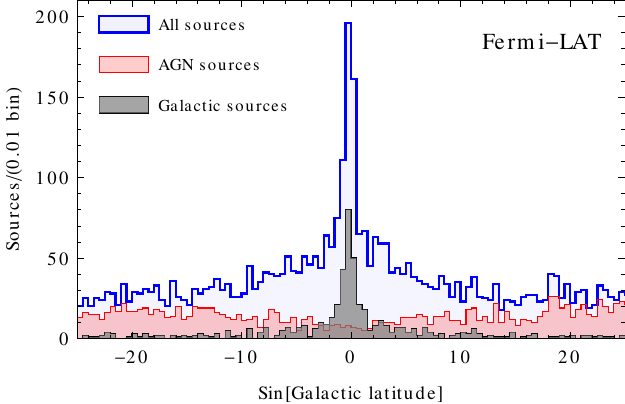}
~~~
\includegraphics[height=4.3cm]{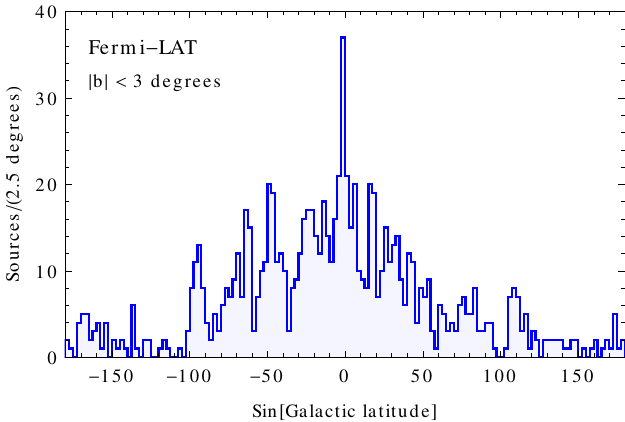}
\end{center}

\vspace{0.05 cm}
\begin{center}
\includegraphics[height=4.3cm]{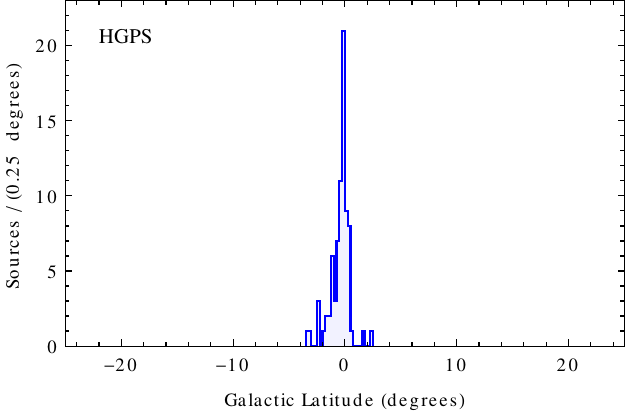}
~~~
\includegraphics[height=4.3cm]{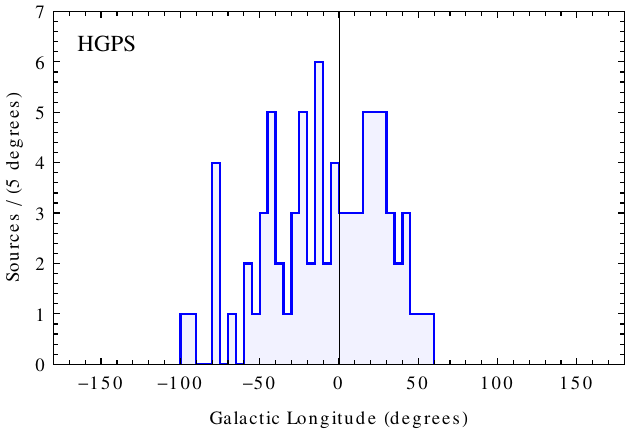}
\end{center}

\vspace{0.05 cm}
\begin{center}
\includegraphics[height=4.3cm]{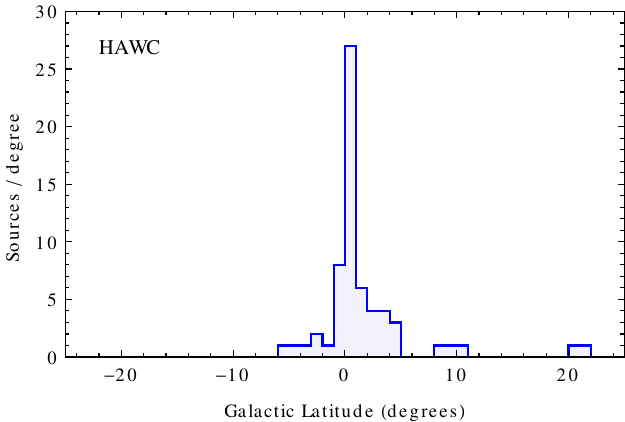}
~~~
\includegraphics[height=4.3cm]{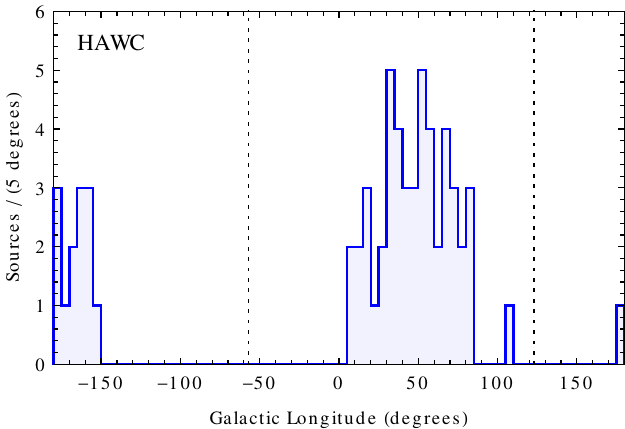}
\end{center}

\vspace{0.05 cm}
\begin{center}
\includegraphics[height=4.3cm]{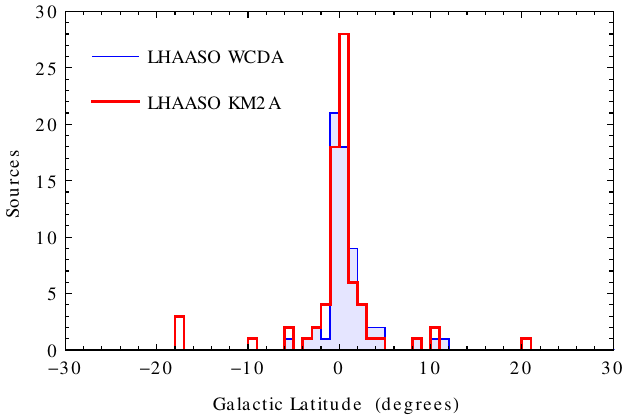}
~~~
\includegraphics[height=4.3cm]{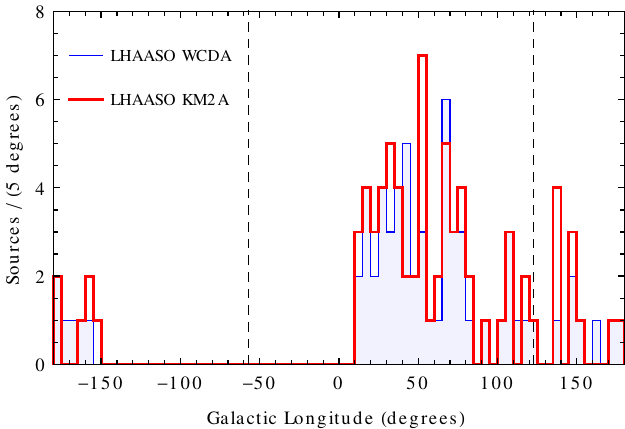}
\end{center}
\caption {\footnotesize Distributions in Galactic latitude and
 longitude for sources in different gamma--ray catalogs.
 Top row for Fermi--LAT \cite{Fermi-LAT-4FGL-dr4}, second row for HESS--HGPS  \cite{HESS:2018pbp},
 third row for HAWC \cite{HAWC:2020hrt},
 and bottom row for LHAASO (WCDA and KM2A) \cite{LHAASO:2023rpg}.
 In the longitude plots for HAWC and LHAASO the vertical dashed lines
 separate the regions in longitude that have equal sensitivity. 
\label{fig:angular_distributions} }
\end{figure}

The sky positions (in Galactic coordinates) of the sources in the two catalogs
are shown in Fig.~\ref{fig:lhaaso_sky}.
It can be seen that most of the sources
are located  close the Galactic equator (the line of latitude $b=0$), 
with distributions that are strongly dependent on the Galactic longitude.
The latitude and longitude distributions of the sources
in the two catalogs are shown in Fig.~\ref{fig:angular_distributions},
together with the distributions for the sources in the
Fermi--LAT \cite{Fermi-LAT:2019yla,Fermi-LAT-4FGL-dr4}
and HESS \cite{HESS:2018pbp} catalogs.

A detailed discussion of the spectral shapes of the Galactic
gamma--ray sources in the catalogs listed above is presented in our companion paper
\cite{lv_spectral_shapes}.

The distributions over the celestial sphere of the objects
listed in the gamma--ray catalogs are determined by the space
and luminosity distributions of the sources,
but also by the sensitivities of the telescopes
and their dependence on direction.
Cherenkov telescopes are pointing instruments
that observe only a small portion of the sky 
at any given time, so the sensitivity of such a telescope 
depends on the history of its observations,
and can generally have a complicated dependence on direction.
The situation is different for an EAS telescope, because its sensitivity,
which  covers a limited but large portion of the sky,
has a dependence on the celestial coordinates that
(to a good approximation for data taking over many sidereal days)
is entirely determined by its geographic latitude, and changes only
gradually and smoothly with direction.
Because of these characteristics of the sensitivity
(discussed in more detail below),
in the present work we use only the observations of 
the EAS telescopes HAWC and LHAASO.

\subsection{Sensitivity of Extensive Air Shower telescopes}
\label{sec:eas-telescopes}
The sensitivity of an EAS gamma--ray telescope,
i.e.  the minimum flux required to
resolve a source is to a good approximation completely determined by the
celestial declination $\delta$ of the direction of observation.
This is because the sky trajectory in local coordinates
of a point on the celestial sphere is determined by the declination $\delta$,
since points of the same declination and different right ascension follow
identical trajectories that differ only by a time shift,
and because the signal from a source must be compared with the
fluctuations of a background dominated by the isotropic cosmic ray flux.

A telescope located at geographic latitude $\lambda$
and  observing only for zenith angle $\theta < \theta_{\rm max}$
can  study the region of the celestial sphere in the declination interval:
\begin{equation}
 \delta \in [\delta_{\rm min}, \delta_{\rm max}]
\label{eq:acc0}
\end{equation}
where the limits: 
\begin{equation}
\begin{array} {l}
 \delta_{\rm min} = {\rm Max}[\lambda - \theta_{\rm max}, -90^\circ], ~ \\[0.15 cm] 
 \delta_{\rm max} = {\rm Min} [\lambda + \theta_{\rm max}, + 90^\circ] 
 \end{array}
\label{eq:acc1}
\end{equation}
are the declinations of points culminating at zenith angle
$\theta= \theta_{\rm max}$ and only touching the cone where measurements
are possible.
Points with declination equal to the latitude of the telescope
($\delta =\lambda$) culminate at the zenith,
and are, to a first approximation, where the sensitivity of the telescope is best.
The sensitivity for any declination can be calculated by integrating over
its trajectory, and taking into account how effective area,
angular resolution, and background rejection depend on the zenith angle.

The fact that the sensitivity of an EAS gammma--ray telescope
is independent of right ascension makes it easier to
compare the observations in two different regions of the sky
if these two regions are chosen to span equal intervals of declination,
and therefore have equal sensitivities.
In this situation, any difference between the results of the observations
for the two regions (e.g. different numbers of resolved sources)
must be attributed to differences in the populations of gamma--ray sources
in the respective Galactic volumes.

Studies of gamma--rays in the Milky Way are best performed using the Galactic
coordinates \{$\ell, b\}$.
The Galactic equator (line of latitude $b=0$) is inclined with respect to the celestial equator
($\delta = 0$), and the two lines intersect at two points
of the celestial sphere that are opposite to each other
with Galactic coordinates $\{\ell,b\} = \{\ell_{1,2}, 0\}$
with longitudes $\ell_1 =32.93^\circ$ and $\ell_2 = -147.07^\circ$.
The two points $\{\ell^*_{\pm,}, 0\}$ on the Galactic equator
with longitudes equidistant from $\ell_1$ and $\ell_2$
(that is with longitudes $\ell^*_+ = 122.93$ and $\ell^*_- = -57.07^\circ$)
are those with the minimum and maximum declination ($\delta = \pm 62.87^\circ$).

This suggests dividing the celestial sphere
into two equal parts by selecting the Galactic longitude ranges
(both extending for $180^\circ$): 
\begin{equation}
\begin{array}{l}
 {\rm Region ~A} : \ell \in [\ell^*_- , \ell^*_+] = [-57.07^\circ , 122.93^\circ] 
\\[0.25cm]
 {\rm Region ~B} : \ell \in
[-180^\circ, \ell^*_-] \oplus [\ell^*_+ , 180^\circ] = 
[-180^\circ, -57.07^\circ]\oplus [122.93^\circ, 180^\circ] 
\end{array}
\label{eq:acc2}
\end{equation}
with the property that the sensitivity of any EAS telescope
is the same in both regions.
This follows from the fact that points on the celestial sphere with longitudes
symmetric with respect to $\ell^*_\pm$, i.e. points with Galactic coordinates 
$\{ \ell^*_{\pm} + \Delta \ell , b\}$ and $\{ \ell^*_{\pm} - \Delta \ell , b\}$
have identical declination.
Note that region~A contains the Galactic Center, and region~B the anticenter.

The two sky regions~A and~B can also be defined 
in terms of celestial coordinates as the right ascension ranges:
\begin{equation}
\begin{array}{l}
 {\rm Region ~A} : \alpha \in [\alpha_1, \alpha_2] = [ -167.14^\circ, +12.86^\circ]
\\[0.25cm]
 {\rm Region ~B} : \alpha \in
 [-180^\circ,\alpha_1] \oplus [\alpha_2, 180^\circ] = 
 [-180^\circ,-167.14^\circ] \oplus [12.86^\circ, 180^\circ]
\end{array}
\label{eq:acc3}
\end{equation}
where the right ascensions $\alpha_1= -167.14^\circ$
and $\alpha_2 = \alpha_2 + 180^\circ$ are the right ascensions of the two points
on a line in the sky of constant declination $\delta$ that have the minimum and maximum
Galactic latitude. In this second definition it is manifest that the
sensitivity of a telescope is equal in both sky regions.
%If an EAS telescope, after a long, continuous period of observations,
%resolves $N_A$ sources in region A and $N_B$ in region B, this implies
%that the populations of sources in the the two sky regions
%are different, without the need to apply corrections.

An EAS telescope, can observe
only a limited portion of the celestial sphere [see Eq.(\ref{eq:acc1})]
and this visible region of the sky can be divided into two parts
of equal solid angle and sensitivity,
one belonging to region~A and the other to region~B.
The visible part of the celestial sphere
for the HAWC and LHAASO telescopes, and their
subdivisions into region~A and region~B are shown in Fig.~\ref{fig:lhaaso_sky}
assuming gamma--ray observations for a maximum zenith angle of
45$^\circ$ for HAWC and 50$^\circ$ for LHAASO.

\begin{figure}
\begin{center}
\includegraphics[width=12.9cm]{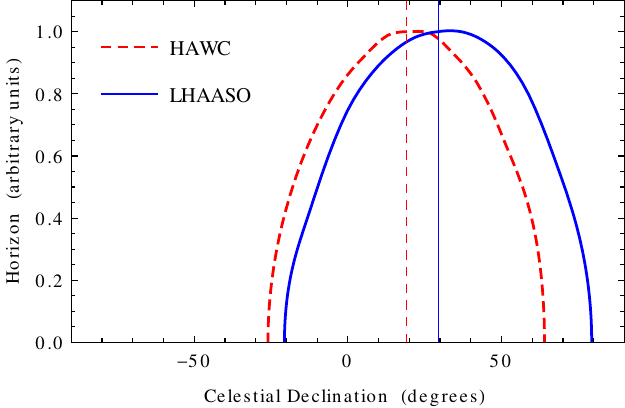}

\vspace{1.25cm}
\includegraphics[width=12.9cm]{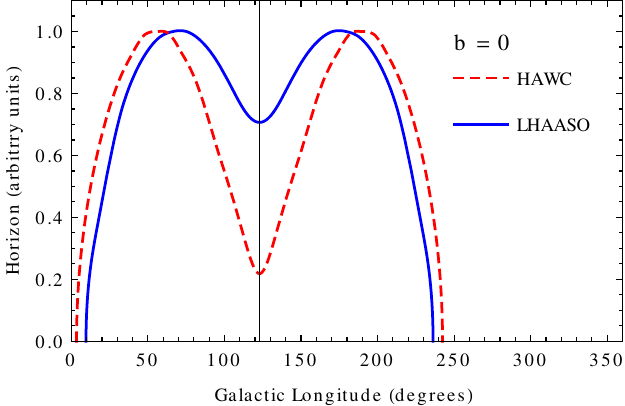}
\end{center}
\caption {\footnotesize
The top panel shows the declination dependence of the horizon to
resolve point--like gamma--ray sources for the
HAWC and LHAASO--KM2A telescopes plotted as a function of celestial declination.
The absolute value of the horizon scales with the source luminosity ($D_H \propto L^{1/2}$).
The horizon for LHAASO--WCDA has a very similar declination dependence. 
The vertical lines correspond to a declination equal to
the geographic latitude of the telescope, where (to a good approximation)
the sensitivity is best, and the horizon is largest.
The bottom panel shows again the horizon 
for points on the Galactic equator ($b = 0$) plotted as a function
of the longitude for the two telescopes.
\label{fig:sens_declination}}
\end{figure}

In \cite{HAWC:2020hrt} and \cite{LHAASO:2023rpg} the HAWC and LHAASO
collaborations give plots of the minimum flux required to resolve a
point--like source as a function of declination. The
declination dependencies of the telescope sensitivities
are shown in Fig.~\ref{fig:sens_declination} re--expressed in terms 
of the relative size of the horizon, that is the maximum distance
to resolve a point--like source. 
The sensitivity of a telescope is best and the horizon is largest
(to a good approximation) 
for $\delta \approx \lambda$ when the source transits through
the local zenith, and the horizon vanishes for
declinations $\delta_{\rm min}$ and $\delta_{\rm max}$ when a source only
grazes the observation cone $\theta < \theta_{\rm max}$.

Since most of the Galactic sources are observed near
the Galactic equator it can be useful to show the
sensitivity as a function of Galactic longitude along
the equator ($b=0$). For the LHAASO detector,
located at longitude $\lambda \simeq 29.36^\circ$,
and assuming an observation cone with $\theta_{\rm max} = 50^\circ$,
the equator is visible only for longitudes
in the range $\ell \in [9.6^\circ, 236.3^\circ]$. 
For the HAWC telescope, located at geographic latitude $\lambda = 19.0^\circ$,
and assuming an observation cone with $\theta_{\rm max} = 45^\circ$,
the Galactic equator is visible in the longitude interval
$\ell \in [3.4^\circ, 242.4^\circ]$.
For both telescopes, for the reasons outlined above,
the dependence of the sensitivity on the longitude 
is symmetrical around the longitude $\ell = \ell^*_+= 122.93^\circ$,
with the part of the sky with $\ell < \ell^*_+$ in region~A ,
and the part of the sky with $\ell > \ell^*_+$ in region~B.

\subsection{Resolved and unresolved gamma--ray sources}
\label{sec:resolved}

A gamma--ray source is completely described by its position, luminosity,
spectral shape and morphology. A precise description of the
spectral shape and morphology of a source can be quite complicated,
and in the following we will make some simplifications.
For the spectral shape, we will assume that, in a not too large energy interval,
it can be reasonably well approximated by a simple power--law form of slope $\alpha_j$
($\phi_j (E) \propto E^{-\alpha_j}$)
[for a discussion of the spectral shapes
of the Galactic gamma ray sources see \cite{lv_spectral_shapes}].

The integrated flux in the interval $[E_{\rm min},E_{\rm max}]$ can then be expressed
(neglecting absorption) in the form:
\begin{equation}
 \Phi_j =
 \frac{L_j}
 {4 \, \pi \, \langle E_j \rangle \; D_j^2 } ~,
\label{eq:phidefint} 
\end{equation}
where $D_j = | \vec{x}_j - \vec{x}_\odot|$ is the distance of the source
from the solar system (located at $\vec{x}_\odot$), and $L_j$ is the source luminosity
in the same energy interval (left implicit in the notation).
The quantity $\langle E_j \rangle$ is the average energy of the photons
in the interval of interest:
\begin{equation}
 \langle E_j \rangle = E_{\rm min} ~\left ( \frac{\alpha_j-1}{\alpha_j -2}
 \right ) \; \left
 [1- \left (\frac{E_{\rm min}}{E_{\rm max}} \right )^{\alpha_j -2} \right ] ~. 
\end{equation}
The morphology of a source will be described 
with the single parameter $R$ that gives its linear size.

A gamma--ray source is resolved in the observations of a telescope
if its integrated flux is larger than a minumum value $\Phi_{\rm min}$
which in general  depends on the direction of observation $\Omega$,
the angular size of the source $\delta\theta_s$ and its spectral shape.
In the following we assume that the sources, in the considered energy interval,
have approximately the same spectral index,
and parameterize the dependence of the minimum flux on the source size
with the form:
\begin{equation}
 \Phi_{\rm min} (\Omega, \delta\theta_s) \simeq \Phi_{\rm min}^0(\Omega) ~
 ~\sqrt{ 1 + \left (\frac{\delta \theta_{s}} {\delta\theta_{\rm exp}}
\right )^2 } ~.
\label{eq:phimin}
\end{equation}
In this expression 
$\Phi_{\rm min}^0 (\Omega)$ is the minimum flux to resolve a
point--like source in the direction $\Omega$,
and $\delta \theta_{\rm exp}$ is the angular resolution of the telescope.
The dependence of the minimum flux on the source angular size
can be derived from the assumption that the minimum observable
signal from a source is proportional to the size of the fluctuations
of an isotropic background (dominated by the hadronic cosmic ray flux)
integrated over a solid angle region of dimension 
$\delta \Omega \propto (\delta \theta_s^2 + \delta\theta_{\rm exp}^2)$.

The condition $\Phi_j > \Phi_{\rm min}$ can be a expressed
as an upper limit on its distance $D_j \le D_H$,
where the horizon $D_H$ is a function of the source luminosity and linear size.
For a point--like object, the horizon (leaving the dependence on the direction $\Omega$ implicit)
takes the simple form:
\begin{equation}
 D_{\rm H}^{0} (L) = \sqrt{\frac{L}
 {4 \pi \, \langle E\rangle \; \Phi_{\rm min}^{0} }} 
\label{eq:thorizon1}
\end{equation}
that grows with the source luminosity $\propto \sqrt{L}$.
If the source has a linear size $R$
the horizon distance is smaller and takes the more general form:
\begin{equation}
 D_{\rm H} (L,R) = D_H^0 (L)
 ~\sqrt{
 \sqrt{1+ \frac{1}{4 \, x^4}} - \frac{1}{2 \, x^2}} 
 \simeq
 \begin{cases}
 D_H^0 (L) ~ x
% = [D_H^{0}(L)]^2 ~(R/\delta\theta_{\rm exp})
 \propto L & {\rm for} ~~x \ll 1 \\[0.25 cm]
 D_H^0 (L) \propto \sqrt{L} & {\rm for} ~~x \gg 1 
\end{cases}
\label{eq:thorizon}
\end{equation}
where the adimensional quantity $x$:
\begin{equation}
 x = \frac{D_H^0(L)}{R /\delta\theta_{\rm exp} }
\label{eq:thorizon2}
\end{equation}
is the ratio between the horizon for a point--like source of luminosity
$L$ and the distance $R/\delta\theta_{\rm exp}$ where a source
of size $R$ has an angular extension equal to the detector resolution.
Note that if the horizon distance is much smaller than $R/\delta\theta_{\rm exp}$,
its dependence on the luminosity is $\propto L$.

%The relation of Eq.(\ref{eq:thorizon}) between luminosity and
%horizon can be inverted to obtain:
%\begin{equation}
%L = L_0 ~\frac{D_H^2}{D_0^2} \, \sqrt{1 + \frac{R^2}{D_H^2}}
%\label{eq:lumfromdist}
%\end{equation}
%where $D_0$ is an arbitrary reference distance, and
%\begin{equation}
%L_0 = 4 \,\pi \, \langle E \rangle \, D_0^2 \; \Phi_{\rm min}^0
%\end{equation}
%is the luminosity of a point--like source that gives the minimum flux
%when at distance $D_0$ from the Earth. 

\subsection{Resolved sources in the HAWC and LHAASO catalogs}
The number of Galactic sources in the HAWC and LHAASO catalogs observed in
regions~A and~B (which are similar but not not identical for the
two telescopes) are listed in Table~\ref{tab:event_numbers}.

\begin{table}[hbt]
\caption{\footnotesize
Numbers of high energy Galactic gamma--ray sources in different sky regions
for the HAWC and LHAASO catalogs.
For the LHAASO catalog the three lines show the sources
detected by the WCDA array, the KM2A array 
and the subset of the KM2A sources detected also above 100~TeV.
Region A corresponds to the Galactic longitude range:
$\ell \le \ell^*$ and region~B to the complementary range
$\ell > \ell^*$, with $\ell^* = 122.9^\circ$.
\label{tab:event_numbers}}
 \renewcommand{\arraystretch}{1.4}

 \vspace{0.20cm}
 \begin{tabular}{ | l || c | c | c | c || c| c | c | c || c | c| c | c | }
 \hline
 ~~~
 & \multicolumn{4}{c||} {All longitudes} 
 & \multicolumn{4}{c||} {$|b| \le 3^\circ$} 
 & \multicolumn{4}{c |} {$|b| > 3^\circ$} \\
 \hline
 Telescope
 & $N$ & $N_A$ & $N_B$ & $N_A/N_B$ 
 & $N$ & $N_A$ & $N_B$ & $N_A /N_B$ 
 & $N$ & $N_A$ & $N_B$ & $N_A /N_B$ \\
 \hline
HAWC &	63	&	49	&	14	&	$	3.5	\pm	1.1	$	&	48	&	46	&	2	&	$	23.	\pm	17.	$	&	15	&	3	&	12	&	$	0.25	\pm	0.16	$	\\
\hline
LHAASO &	86	&	66	&	20	&	$	3.3	\pm	0.84	$	&	70	&	61	&	9	&	$	6.8	\pm	2.4	$	&	16	&	5	&	11	&	$	0.45	\pm	0.25	$	\\
\hline
LHAASO--WCDA &	65	&	51	&	14	&	$	3.6	\pm	1.1	$	&	55	&	49	&	6	&	$	8.2	\pm	3.5	$	&	10	&	2	&	8	&	$	0.25	\pm	0.2	$	\\
 \hline
LHAASO--KM2A &	75	&	58	&	17	&	$	3.4	\pm	0.94	$	&	62	&	54	&	8	&	$	6.8	\pm	2.6	$	&	13	&	4	&	9	&	$	0.44	\pm	0.27	$	\\
 \hline
LHAASO--KM2A-High~ &	44	&	36	&	8	&	$	4.5	\pm	1.8	$	&	34	&	32	&	2	&	$	16.	\pm	12.	$	&	10	&	4	&	6	&	$	0.67	\pm	0.43	$	\\
 \hline
\end{tabular}
\end{table}

In the case of LHAASO we show the results for the measurements made 
by the WCDA and KM2A arrays, and also the results for the KM2A sources that have
measurable fluxes for energies $E \ge 100$~TeV.
When looking at these numbers one should keep in mind that the
sensitivity of the telescopes in the A and B regions is identical.
The number of sources
$\{N_A, N_B\}$ for the four cases are:
\{49,14\}, 
\{51,14\}, 
\{68,17\} and \{36,8\} (for HAWC, LHAASO--WCDA,
LHAASO--KM2A and LHAASO--high respectively)
with $N_A/N_B$ ratios  between 3.5 and 4.5.
The numbers of resolved sources are small,
but it is clear that there is a large asymmetry,
with more sources in the sky region closer to the Galactic Center.
This suggests that the sources can be resolved at distances larger than a few kiloparsecs.
Such a conclusion is clearly important for estimating the flux from the ensemble
of unresolved sources, located beyond this horizon.
In the following we will try to obtain a more precise estimates
for the horizon and the unresolved flux.

Table~\ref{tab:event_numbers} also lists the number of sources
in two different ranges of latitude ($|b| \le 3^\circ$ and $|b| > 3^\circ$).
Most of the sources are observed at small $|b|$, and 
the ratio $N(|b| \le 3^\circ)/N({\rm all}~b)$
is 0.76 for HAWC and 0.81 for LHAASO. 
This indicates that the sources are distributed in a thin disk,
and that the horizon is much larger than the thickness of this disk.

Table~\ref{tab:event_numbers} also gives the number of resolved sources
of small ($|b| \le 3^\circ$) and large ($|b| > 3^\circ$) latitude
in the A~and~B sky regions. These numbers reveal the surprising fact
that for sources at large latitude one observes 
more sources in region~B, toward the Galactic anti--center.
The number of sources $\{N_A(|b| < 3^\circ), N_B(|b| < 3^\circ)\}$
in the same four catalogs as before are: \{3,12\}, \{2,8\}, 
\{4,9\} and \{4,6\}, and the ratio $N_A/N_B$ is significantly less than unity.
This result, that is also clearly visible 
in Fig.\ref{fig:lhaaso_sky} where it can be seen that the
latitude spread for sources at large $|\ell|$
is larger  than for sources closer to the Galactic Center.
This result is due to the fact that the sources observed at large $|b|$
are near  the Solar system, and their space distribution is determined
by the position and shape of the spiral arms of the Galaxy.

This result suggests that smooth, cylindrically symmetric models of
the space distributions of the sources are not adequate to describe these
features, and that a precise and detailed modeling of the spiral arms is required.
For this reason (and because of the small statistics available
at the moment) we postpone the study of the declination
dependence of the gamma--ray sources
to a future work, and in the following
we will base our discussion on the comparison of only two large sky regions,
because in this case the sensitivity to the precise modeling of the
the space distributions of the sources is much reduced.

\section{Diffuse Galactic gamma--ray flux measurements}
\label{sec:diffuse}
In the the energy range 0.1--100~GeV the flux
generated by the emission from cosmic rays propagating in
the interstellar medium has been well measured
over the entire celestial sphere by
the Fermi--LAT telescope \cite{Fermi-LAT:2012edv,Fermi-LAT:2016zaq}.
Below few GeV the emission generated by relativistic electrons
and positrons (via Compton scattering
and bremsstrahlung) is important.
At higher energies most of the diffuse flux is
generated by the hadronic mechanism,
where gamma--rays are created in the decay of
mesons (mostly $\pi^\circ$) produced in the inelastic collisions of
protons and nuclei with interstellar gas.
The hadronic mechanism is also expected to be dominant in the TeV
and PeV energy range.

The Fermi--LAT observations of the diffuse flux
encode information that is very important for high energy astrophysics.
The crucial problem is to determine whether or not cosmic rays
in different parts of the Milky Way have spectra of the same shape
as  measured in the vicinity of the Earth.
There is currently some controversy about the interpretation of the Fermi--LAT measurements
(see for example \cite{Gaggero:2014xla,Yang:2016jda}), and 
observations at higher energies are of crucial importance to
shed light on the problem.

Measurements of the diffuse gamma--ray flux for $E \gtrsim 1$~TeV are
possible only with telescopes at ground level, and therefore
measurements have been obtained only in some limited regions of the sky.

\begin{itemize}
\item The ARGO--YBJ telescope \cite{ARGO-YBJ:2015cpa} has published a
measurement of the diffuse flux in the (approximate) energy range [0.35,2]~TeV
for the sky region of Galactic latitude $|b| < 5^\circ$ and longitude
$25^\circ < \ell < 100^\circ$.

\item The Tibet--AS$\gamma$ telescope \cite{TibetASgamma:2021tpz} has published
measurements of the diffuse flux in the (approximate)
energy range [10$^2$,10$^3$]~TeV in two sky regions (in partial overlap)
\{$|b| < 5^\circ$, $25^\circ \le \ell \le 100^\circ$\} and
\{$|b| < 5^\circ$, $50^\circ \le \ell \le 200^\circ$\}.
(note that the first region in the same as that used by ARGO--YBJ).

\item The LHAASO--KM2A detector \cite{LHAASO:2023gne}
has published measurements of the diffuse
gamma--ray flux in the two regions of the sky around the Galactic plane:
\{$|b| < 5^\circ$, $15^\circ \le \ell \le 125^\circ$\} and
\{$|b| < 5^\circ$, $125^\circ \le \ell \le 235^\circ$\}.
These two regions have the same solid angle extent,  with the second 
(Outer--Galaxy) centered on the Galactic anticenter and the other
(Inner--Galaxy) approaching but not reaching the Galactic center.

\item The HAWC telescope \cite{HAWC:2023wdq} has also published
a measurement of the diffuse gamma--ray flux in the smaller sky region
\{$|b| < 4^\circ$, $43^\circ \le \ell \le 73^\circ$\}
The HAWC publication also includes flux estimates  in  smaller portions
of this region.
\end{itemize}

\begin{figure}
\begin{center}
\includegraphics[width=14cm]{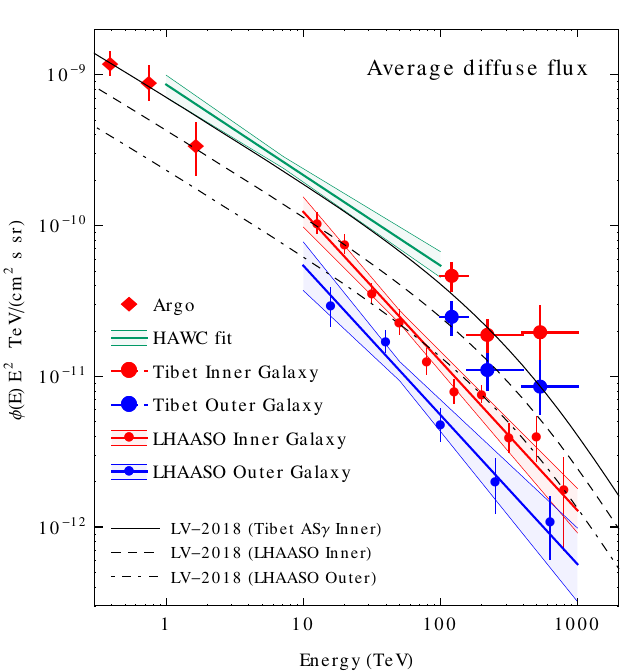}
\end{center}
\caption {\footnotesize
 Measurements of the average diffuse gammma--ray flux at high energy
 obtained by different telescopes:
 ARGO-YBJ \cite{ARGO-YBJ:2015cpa}, Tibet~AS$\gamma$ \cite{TibetASgamma:2021tpz},
 HAWC \cite{HAWC:2023wdq}  and LHAASO \cite{LHAASO:2023gne} in different sky regions.
 The predictions of the average diffuse flux for the
 factorized model in LV-2018 \cite{Lipari:2018gzn} for the same sky regions
 are also shown in the figure.
 \label{fig:diffuse_all_data}}
\end{figure}

The diffuse flux measurements of ARGO, Tibet--As$\gamma$, LHAASO and
HAWC are shown in Fig.~\ref{fig:diffuse_all_data}
in the form of the average flux in the sky region of observation.
It should be noted that the measurements are made
in different regions of the sky and therefore a comparison
of the results requires a careful discussion of the angular dependence
of the flux.

In Fig.~\ref{fig:diffuse_all_data} we also show (as thin lines) the predictions
of the diffuse flux of the factorized model of reference
\cite{Lipari:2018gzn} (hereafter LV--2018 ). This model assumes
that the angular distribution of the diffuse flux is energy independent
and remains equal to what has been observed by Fermi--LAT
for energy $E \simeq 10$~GeV, and the energy dependence of
the flux obtained assuming that for $E\gtrsim 10$~GeV the emission is 
dominated by the hadronic mechanism (see below)
and can therefore be calculated, and that
the shape of the cosmic ray spectra in all points of the Galacy
is identical to what is observed in the vicinity of the Earth.
Uncertainties in the shape of the energy spectrum are then 
controlled by the modeling of the observed cosmic ray spectra
and composition in the range where only indirect EAS measurements are possible.
The resulting gamma--ray spectrum is then in first
approximation a power law with a spectral index $\alpha \approx 2.7$,
which gradually softens for $E \gtrsim 10^2$~TeV, due to the
``Knee'' in the primary particle spectrum.

Table~\ref{tab:diffuse_angular} shows the relative magnitude  of the
average diffuse flux observed by Fermi--LAT in  different sky regions
at $E \simeq 12$~GeV, when the most of the flux
is produced  by the hadronic mechanism.
%%%%%%%%%%%%%%%%%%%%%%%%%%%%%%%%%%%%%%%%%%%%%%%%%%%%%%%%%%%%%%%%%%%%%%%%%%%%%%%%%%
\begin{table}[hbt]
 \caption{\footnotesize
 Relative magnitude of the average diffuse (interstellar emission) gamma--ray flux
 estimated from the Fermi--LAT observations at $E=12$~GeV in the sky regions
 where measurements at high energies by Tibet--AS$\gamma$ \cite{TibetASgamma:2021tpz},
 LHAASO \cite{LHAASO:2023gne} and HAWC \cite{HAWC:2023wdq} are available.
 For the two LHAASO regions the average flux is also calculated in the solid angle
 remaining after the masking of known sources.
\label{tab:diffuse_angular}}
\renewcommand{\arraystretch}{1.4}

\vspace{0.2 cm}
\begin{tabular}{ | l | c | c || c || c | c | c |}
 \hline
 Sky region & Latitude & Longitude & ~~ $\langle \phi \rangle_{\rm no-mask}$ ~~ &
$\Delta \Omega_{\rm mask} / \Delta \Omega $ & 
 $\langle \phi_{\rm mask} \rangle / \langle \phi_{\rm no~mask} \rangle$
 & ~~ $\langle \phi \rangle_{\rm mask}$ ~~ \\
 \hline
LHAASO: Inner Galaxy & $|b| < 5^\circ$ & $ 15^\circ \le \ell \le 125^\circ$ & 1 & 0.617 & 0.622 & 1 \\
LHAASO: Outer Galaxy & $|b| < 5^\circ$ & $125^\circ \le \ell \le 235^\circ$ & 0.337 & 0.802 & 0.984 & 0.534 \\
HAWC: Innner Galaxy & $|b| < 4^\circ$ & $43^\circ \le \ell \le 73^\circ$ & 0.960 & -- & -- & 1.54 \\
Tibet--AS$\gamma$: Inner Galaxy ~ & $|b| < 5^\circ$ & $25^\circ \le \ell \le 100^\circ$ & 1.027 & -- & -- & 1.65 \\
Tibet--AS$\gamma$: Outer Galaxy ~ & $|b| < 5^\circ$ & $50^\circ \le \ell \le 200^\circ$ & 0.530 & -- & -- & 0.852 \\
\hline
\end{tabular}
\end{table}
From the table one can see that in three regions
(Tibet--AS$\gamma$ Inner Galaxy, LHAASO Inner--Galaxy, and HAWC)
the average fluxes are approximately equal 
with ratios (1:0.97:0.93),
while the Tibet--AS$\gamma$ Outer--Galaxy region has
a flux that is a factor 0.53 smaller, and the LHAASO Outer--Galaxy region
a fraction 0.33 smaller.
It is important to note, however, that the LHAASO 
diffuse flux measurements are made after masking the points in the sky around
resolved gamma--ray sources, and therefore exclude
38.3\% and 19.8\% of the Inner--Galaxy and Outer--Galaxy regions, respectively.
The gamma--ray sources are mostly located at small Galactic latitudes,
and therefore the masking mostly excludes  points at  small $|b|$,
where also the flux due to the  interstellar emission is higher.
Using the results of Fermi--LAT to model the angular  distribution of the
diffuse flux, one can estimate that the average  fluxes
in the solid angle reduced  by the masking 
are lower by factors of 0.622 and 0.984
for the Inner and Outer--Galaxy regions, respectively.

One should keep in mind that what is observed as a diffuse gamma--ray
flux can be considered, at least in first approximation, as formed by the sum
of a (truly diffusive) component generated by
the interstellar space emission, and the flux due to the ensemble of all
unresolved sources:
\begin{equation}
\Phi_{\rm diffuse}^{\rm data} =
\Phi_{\rm ism} + \Phi^{\rm unresolved }_{\rm sources} ~.
\label{eq:flux_diffuse_composition}
\end{equation}
In this expression the fluxes are integrated in energy
over the interval $[E_{\rm min},E_{\rm max}]$ and in angle
over the sky region $\Delta \Omega$, that are both left implicit.
Disentangling these components is of great importance.

A comparison of the diffuse flux observations by LHAASO and
Tibet--AS$\gamma$ will be briefly discussed below in section~\ref{sec:nu_icecube}.

\section{The Horizon method} 
\label{sec:method}
The main goal of the present work is to estimate luminosity of the ensemble
of gamma--ray sources in the Galaxy, and the flux of unresolved sources
extrapolating from the information obtained from the
observations of the resolved sources.
To perform this task we will use a simple method
based on the study of the distribution in celestial coordinates
of the sources contained in the catalogs of the EAS telescopes. 

To illustrate the idea behind this method one can
consider a simple, ideal situation where the Milky Way contains
$N_{\rm tot}$ identical point--like sources of luminosity $L_0$,
and a telescope observes a region $\Delta \Omega$ of the celestial sphere,
with a sensitivity independent of the direction, and therefore 
resolving all sources within the horizon distance $D_H$, that is a function of
the source luminosity according to Eq.(\ref{eq:thorizon1}).

We will now assume that the space distribution of the gamma--ray sources
is known. Of course, this space distribution
should be determined using present and
future observations, however, it is reasonably safe to assume that the gamma--ray sources
have a distribution similar to other components of the Milky Way
(such as stars, pulsars or supernova remnants) and form a thin disk that
extends for several kiloparsecs around the Galactic center,
with the Solar system placed at a radius of approximately 8.5~kpc. 

The result that the telescope has resolved $N_s$ sources does not
allow to infer the values of $L_0$ and $N_{\rm tot}$
because the observations can be reproduced by
an infinite number of solutions.
In some solutions the Galaxy contains many
faint sources, with the telescope resolving only close objects,
in other solutions the Galaxy contains a smaller number of
brighter objects that can be resolved in a larger volume.

To resolve this ambiguity one can divide the sky into two parts:
region~A toward the Galactic Center and region~B
toward the anti-center, that contain $N_A$ and $N_B$ resolved sources.
It is not strictly necessary, but it is convenient
to choose the two regions as having equal solid angle ($\Delta \Omega_A = \Delta\Omega_B$).
It is now easy to see that the pair of values $\{N_A, N_B\}$ can be mapped into
the pair of quantities $\{L_0, N_{\rm tot}\}$ (and viceversa).

In fact the ratio $N_A/N_B$ depends only on the size of the horizon $D_H$
and therefore only on $L_0$ (the luminosity of the individual sources).
Since the source density increases toward the Galactic center,
a ratio $N_A/N_B$ of the order of unity indicates that all resolved sources
are near the Solar system, and therefore that the horizon $D_H$ is short
and the luminosity $L_0$ is small.
Increasing the horizon $D_H$ (and therefore the luminosity $L_0$)
the ratio $N_A/N_B$ grows because the volume
subtended by the solid angle $\Delta \Omega_A$
(toward the Galactic Center) and of radius $D_H$ 
contains a larger fraction of the Milky Way than the
corresponding volume subtended by the solid angle $\Delta \Omega_B$
(toward the anti--center).
The ratio $N_A/N_B$ reaches a maximum asymptotic value for very large $L_0$,
when the horizon $D_H$ becomes larger than the maximum extension of the Galaxy
along any direction (that is for $D_H \gtrsim 50$~kpc)
and all Galactic gamma--ray sources in the field of view
of the telescope are resolved.

Having estimated $L_0$ from the ratio $N_A/N_B$,
it then straightforward to infer the total number
of sources and the total luminosity of the Milky Way from the
number of resolved sources $(N_A + N_B)$.
It is also obvious that having determined the horizon beyond which the sources
are not resolved, one can make use of the (assumed) space distribution of the sources,
to compute the flux of unresolved sources in any desired direction.

\begin{figure}

 \begin{center}
 \includegraphics[height=5.5cm]{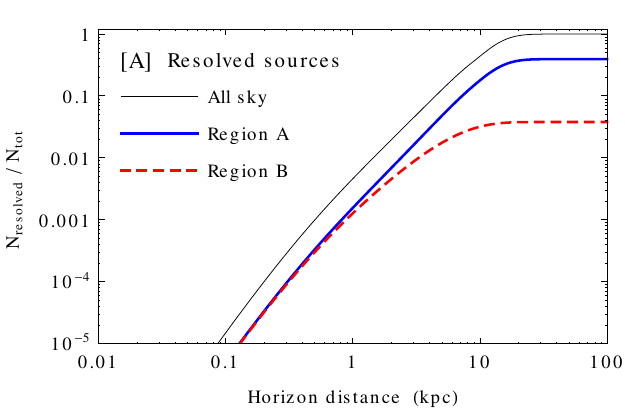}
 ~~~
 \includegraphics[height=5.5cm]{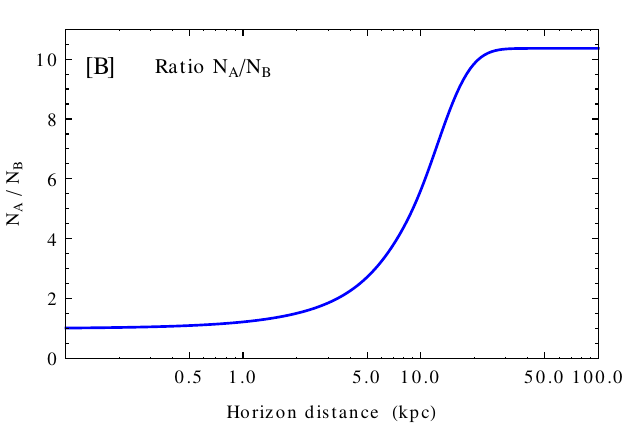}
\end{center}

\vspace{0.2 cm}
\begin{center}
 \includegraphics[height=5.5cm]{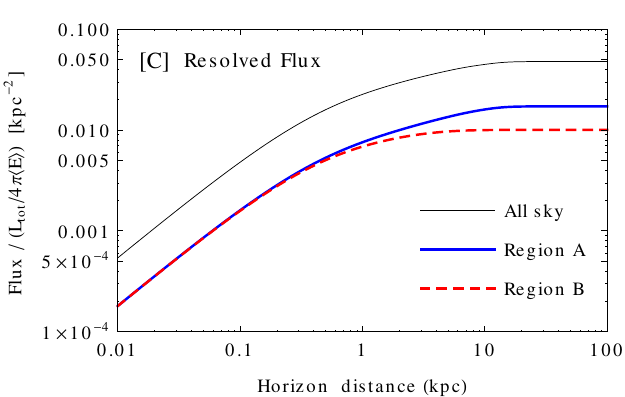}
~~~
 \includegraphics[height=5.5cm]{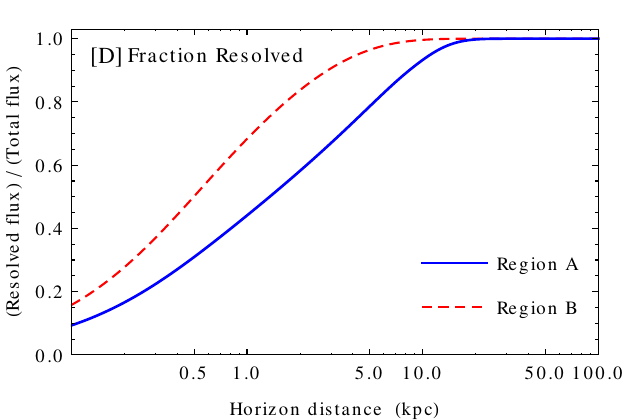}
\end{center}

\vspace{0.2 cm}
\begin{center}
 \includegraphics[height=5.5cm]{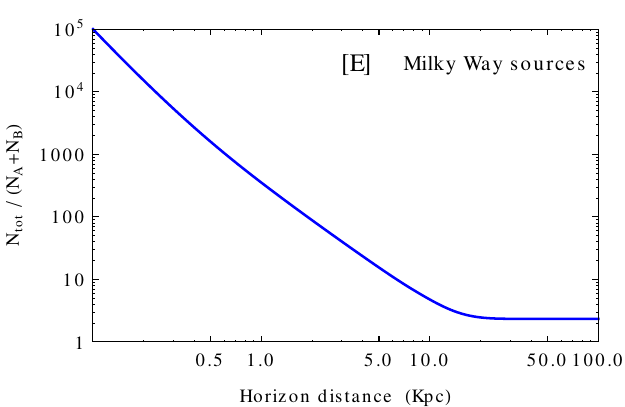}
 ~~~
 \includegraphics[height=5.5cm]{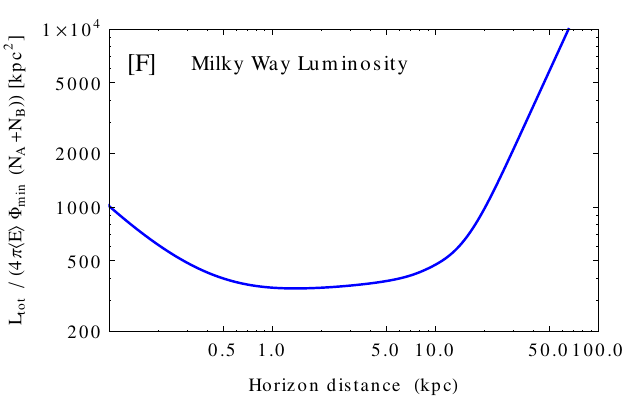}
\end{center}

\caption {\footnotesize 
This figure illustrates how the ``horizon method'' allows to estimate:
the luminosity $L_0$ of the individual gamma--ray sources, 
the total luminosity $L_{\rm tot}$ of all Milky Way sources, and the
flux of unresolved sources on different sky regions.
In this example we assume that the Galactic sources are all point--like
and identical, and that the minimum flux $\Phi_{\rm min}$ to resolve a source
is direction independent. The calculation is performed for the
LHAASO telescope, using the cylindrically symmetric model
for the space distribution of the sources.
The six panels show different quantities plotted as a
function of the horizon distance that is in  one--to--one correspondence
with the luminosity $L_0$
[see Eq.(\ref{eq:thorizon1})].
 Panel [A] shows the number of resolved sources 
 for all the sky, and for regions~A and~B.
 Panel [B] shows the ratio $N_A/N_B$ of the numbers of resolved sources in the two regions.
 Panel [C] shows the flux of resolved sources in all sky and
 in regions~A and~B divided by the quantity $L_{\rm tot}/(4\pi \, \langle E\rangle)$
 (with $\langle E\rangle$ the average energy of photons on the interval considered).
 Panel [D] shows the fraction of the source flux
 due to resolved sources in region~A and region~B. Panel [E] shows the
 ratio $N_{\rm tot}/(N_A + N_B)$ between the total number of sources
 in the Milky Way and the number of resolved sources.
 Panel [F] shows the ratio of the total luminosity
 of the Milky Way and the quantity
 ($4 \pi \, \langle E\rangle \, \Phi_{\rm min} \, (N_A + N_B)$).
 \label{fig:geometry_sources}}
\end{figure}

The different steps of the horizon method are shown graphically in the six panels
of Fig.~\ref{fig:geometry_sources} where different quantities associated with the properties
of the Galactic gamma--ray sources are plotted as a function of the horizon distance.
This example is calculated assuming
the cylindrically symmetric model for the space distribution of the gamma--ray sources
discussed in the appendix~\ref{sec:cylindrical}, 
the simple model of point--like identical sources and a direction
independent minimum flux $\Phi_{\rm min}$ (in the sky region visible by the LHAASO telescope).
It is simple to see how the measurement of $N_A$ and $N_B$ allows to determine
$L_0$, $L_{\rm tot}$ and the resolved and unresolved fluxes in the two sky regions~A and~B.
Note in panel~F how the estimate of the total Galactic luminosity is
approximately constant for the horizon in the interval 1--10~TeV. 

\subsection{Modeling of the Galactic sources}
In the following we will apply the simple geometrical idea introduced above to interpret
the observations of the HAWC and LHAASO telescopes.
As already discussed, the optimal way to divide the portion of the
celestial sphere observed by an EAS telescopes into two parts is to
use the subdivision descrived by Eq.(\ref{eq:acc2}) or equivalently Eq.(\ref{eq:acc3}).
The two regions obtained in this way are of equal solid angle,
and have (in very good approximation)
identical sensitivities, therefore any difference in the number of observed sources
must be attributed to real differences in the distributions of the gamma--ray sources.

We want to perform a reasonably realistic calculation,
however we will have to introduce some simplifications:
\begin{enumerate}
\item We will assume that all sources have the same spectral shape: a power--law
 of slope $\alpha$. Our discussion will always refer to the
 flux and luminosity in rather narrow energy intervals,
 so that this will be a reasonable approximation.
\item We will also assume that all sources have the same linear size $R$,
 and then study how the results depend on the value of $R$.
\item We will assume the factorization of the space and luminosity dependence
 of the source distributions:
\begin{equation}
\frac{dN_s}{d^3 x \, dL} = f(\vec{x}) \, \frac{dN_s}{dL}
\label{eq:factorized}
\end{equation}
This factorization hypothesis is motivated by the search for simplicity,
and should be critically reviewed in future studies.
\end{enumerate}
In Eq.(\ref{eq:factorized}) the function $f(\vec{x})$ describes the space distribution
of the sources and, without any loss of generality,
is normalized to unity:
\begin{equation}
\int d^3 x ~f(\vec{x}) = 1 ~.
\end{equation}
In the following we will use a model with cylindrical symmetry,
based on previous works of Yusifov and Kucuk \cite{Yusifov:2004fr}
that fit the distribution of pulsars in the Galaxy, and
a model that includes the spiral structure of the Milky Way based on the studies
of Reid et al. \cite{Reid:2019} that fit the 
 space distribution of high--mass star forming regions.
Both models are presented in the appendix~\ref{sec:space_distributions}.

\begin{figure}
\begin{center}
\includegraphics[width=14cm]{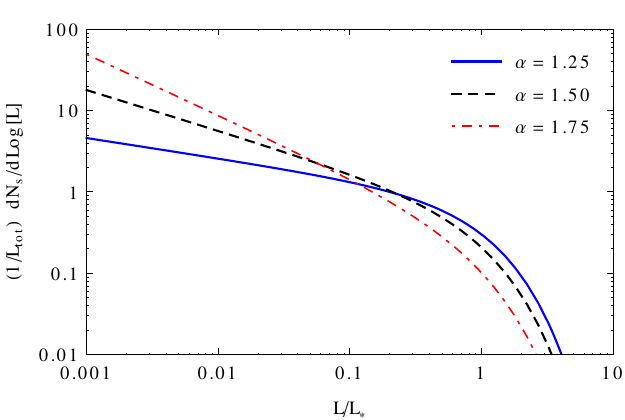}

\vspace{0.25cm}
\includegraphics[width=14cm]{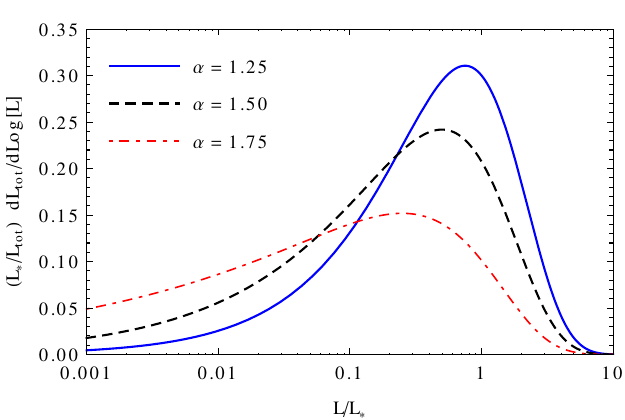}
\end{center}
\caption {\footnotesize
 The top panel shows the luminosity distribution 
 with the form of Eq.~(\ref{eq:lum_dist}) (a power--law with exponential cutoff),
 plotted in the form $L \, dN_s/dL = dN_s/d\log L$
 for three values of the exponent: $\gamma = 1.25, 1.5$ and 1.75.
 In the absence of a low luminosity cutoff the total number of sources diverge.
 The bottom panel shows again the luminosity distribution
 (for the same three values of the exponent)
 in the form $L^2 \, dN_s/dL = dL_{\rm tot}/d\log L$ plotted with a linear $y$ scale.
 The total luminosity remains finite.
 \label{fig:lum_dist}}
\end{figure}

For the luminosity distribution of the sources we will
consider the simple toy model where all sources are identical
with luminosity $L_0$:
\begin{equation}
\frac{dN_s}{dL} = \frac{L_{\rm tot}}{L_0} ~\delta [L - L_0]
\label{eq:lum0_dist}
\end{equation}
and a more realistic model, a power law with exponential cutoff:
\begin{equation}
\frac{dN_s}{dL} = \frac{L_{\rm tot}}{ \Gamma (2 -\gamma) \; L_*^2}
~\left ( \frac{L}{L_*} \right )^{-\gamma}
~\exp \left[ - \frac{L}{L_*} \right ] ~.
 \label{eq:lum_dist}
\end{equation}
This distribution is determined by the three parameters
\{$L_{\rm tot}$, $L_*$, $\gamma\}$,
with $L_{\rm tot}$ the total luminosity
of the ensemble of all Galactic sources,
$L_*$ a characteristic luminosity, and the exponent $\gamma$ 
controls the importance of faint sources.
A luminosity distribution of very similar form 
(a power-law between sharp limits $L_{\rm min}$ and $L_{\rm max}$)
was introduced by Strong in \cite{Strong:2006hf},
and has subsequently been used by other authors \cite{Steppa:2020qwe,Cataldo:2020qla}.
The form of Eq.(\ref{eq:lum_dist}) does not have a low luminosity cutoff,
and replaces the sharp high energy cutoff with a more realistic exponential one.
The lack of the low energy cutoff implies that the
total number of sources in the Galaxy diverges for $\gamma \ge 1$,
while the total luminosity diverges for $\gamma \ge 2$.
An illustration of luminosity distribution in Eq.(\ref{eq:lum_dist})
is given in Fig.~\ref{fig:lum_dist}, where one can see that,
for an exponent $\gamma$ smaller and not to close to two,
the total luminosity of the Galactic sources,
and therefore the total source flux, is mostly
due to objects that have a luminosity around $L_*$.

It could appear surprising that the absolute normalization of the
luminosity distribution is parametrized with $L_{\rm tot}$,
the total power of the Milky Way
gamma--rays sources, and not with $N_{\rm tot}$, the total number of sources.
This is however a better choice, because the
total number of sources in the Galaxy
is a quantity that can be ambiguous and misleading, and
is in fact essentially impossible to measure, because 
the Galaxy very likely contains a large number of very faint sources
that give a negligible contribution to both the total flux
and the total luminosity. In fact, as already discussed,
the total number of sources in the Galaxy diverges for an exponent
$\gamma \ge 1$, but this divergence is harmless, as long as the total
luminosity of the Galaxy remains finite.

The number of sources resolved by a telescope
in the sky region $\Delta \Omega$ can be calculated
integrating over the luminosity and space distributions of the source:
\begin{equation}
 N_s (\Delta \Omega) =
\int dL ~\frac{dN_s} {dL} 
~ \int_{\Delta \Omega} d\Omega ~\int_0^{D_H (L, \Omega)} dt~t^2
~f [ \vec{x}_\odot + \hat{n}(\Omega) \, t] 
\label{eq:n_sources}
\end{equation}
where the integration in space is extended up to the horizon $D_H(L, \Omega)$
that depends on the luminosity of the source
and on the sensitivity of the telescope in that direction
[see Eq.(\ref{eq:thorizon})].

The total flux from all (resolved and unresolved) gamma--ray sources
in the sky region $\Delta \Omega$ can be calculated as:
\begin{equation}
 \Phi_{\rm sources}^{\rm all} (\Delta \Omega) = \frac{L_{\rm tot}}{4 \, \pi \; \langle E \rangle } ~
 \int_{\Delta \Omega} d\Omega ~\int_0^{\infty} dt
~f [ \vec{x}_\odot + \hat{n}(\Omega) \, t] 
\label{eq:flux_all}
\end{equation}
and is proportional to the total Galactic luminosity and is 
independent of the form of the luminosity distribution.
The flux from unresolved sources in the same sky region
does depend on the luminosity distribution and can be obtained
with an integration similar to the one in Eq.(\ref{eq:n_sources}):
\begin{equation}
 \Phi_{\rm sources}^{\rm unresolved} (\Delta \Omega) = \frac{1}{4 \, \pi \; \langle E \rangle } ~
\int dL ~L ~\frac{dN_s} {dL} 
~ \int_{\Delta \Omega} d\Omega ~\int_{D_H (L, \Omega)}^\infty dt
~f [ \vec{x}_\odot + \hat{n}(\Omega) \, t] 
\label{eq:flux_unr}
\end{equation}
where the space integration extends from the horizon to infinity.

For a luminosity distribution of the form of Eq.~(\ref{eq:lum_dist}),
the sky distribution of the resolved sources can be used to
estimate the parameters $L_*$ and $L_{\rm tot}$.
Assuming the validity of the factorization hypothesis,
a form for the space distributions of the sources,
and a value for the exponent $\gamma$, then it is straightforward to
map the pair $\{N_A, N_B\}$
(the numbers of resolved sources in region~A and~B)
into the parameters $\{L_*,L_{\rm tot}\}$ following the same method
outlined above for the case of identical sources.
The ratio $N_A/N_B$ is a function only of $L_*$,
and then the absolute number of resolved sources
allows to estimate $L_{\rm tot}$.

It should be noted that for the power--law luminosity model,
if the exponent is $\gamma \ge 1$
(and the total number of sources in the Galaxy diverges),
in the limit $L_* \to \infty$, the number of resolved also diverges,
as more and more faint sources can be identified.
The rate of the divergence of the number of resolved sources
depends on the exponent $\gamma$ and it is faster for larger
exponents with the form:
$N_{\rm resolved} \propto L_*^{(\gamma-1)}$.

\begin{figure}
\begin{center}
\includegraphics[width=12.5cm]{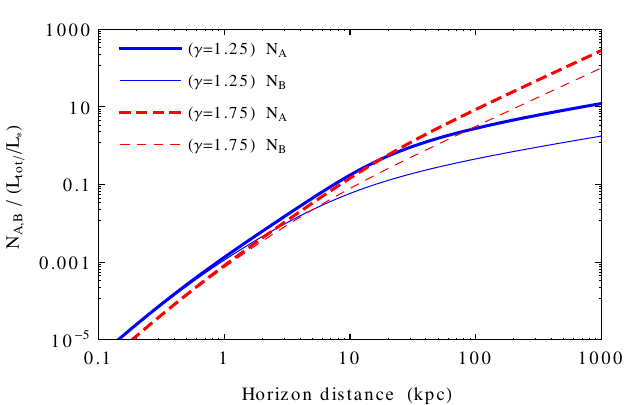}
\end{center}
\caption {\footnotesize
Number of resolved sources in region~A and~B of the LHAASO detector
calculated assuming  the cylindrically symmetric space distribution described in
appendix~\ref{sec:cylindrical} and  the power--law with exponential cutoff for
the luminosity distribution [see Eq.(\ref{eq:lum_dist})].
The number of sources is plotted as a function
of the horizon for the critical luminosity $L_*$ in a direction of declination
$\delta = \lambda$. The curves are calculated
for two values of the exponent of the power law: $\gamma = 1.25$ and $\gamma = 1.75$.
For large values of $L_*$, when the horizon is larger than
the linear extension of the Galaxy, the number of resolved sources
increase as a power law $N_{A,B} \propto L_*^{\gamma-1} \propto D_H^{2(\gamma-1)}$.
Note also how the ratio $N_A/N_B$ for a large horizons
is smaller for the larger exponent $\gamma$.
 \label{fig:geom_n_plaw}}
\end{figure}

The divergence can be also expressed in terms of the horizon
$D_*= D_H (L_*, \Omega)$ for a source of luminosity $L_*$ in an
fixed (arbitrary) direction: $N_{\rm resolved} \propto D_*^{2 (\gamma-1)}$.
This is illustrated in Fig.~\ref{fig:geom_n_plaw}
that plots, for the LHAASO detector, 
the number of resolved sources in regions~A and~B
as a function of the horizon $D_* = D_H (L_*, \delta = \lambda)$
for the critical luminosity $L_*$ in a direction of declination 
$\delta =\lambda$ for two values of the exponent $\gamma$.
For large $L_*$ the numbers of resolved sources in different angular regions
grow at the same rate, and therefore 
the ratio $N_A/N_B$ becomes asymptotically a constant,
that depends on the exponent $\gamma$.

\begin{figure}
\begin{center}
\includegraphics[width=12.5cm]{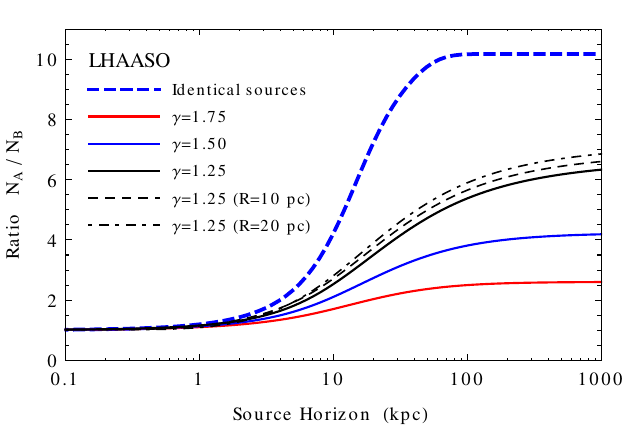}
\end{center}
\caption {\footnotesize
 The top panel shows how the ratio $N_A/N_B$ of gamma--ray
 sources observable in region~A and region~B for the LHAASO telescope
 depends on the horizon for source resolution, in the case of identical
 sources and for the luminosity distribution of
 Eq.~(\ref{eq:lum_dist}) (a power--law with exponential cutoff).
 The horizon indicated in the $x$--axis is for a source 
 with celestial declination $\delta=\lambda$
 (with $\lambda$ the geographic latitude of the telescope).
 For other directions the horizon is shorter, and depends on
 $\delta$ with the form shown in Fig.~\ref{fig:sens_declination}.
 One line is for point--like identical sources, the other lines
 are for a power--law luminosity distribution, and in this case the
 horizon indicated in the $x$--axis is for sources
 with the cutoff luminosity $L_*$,
 and scales with $L$ according to Eq.~(\ref{eq:thorizon}).
 Three curves are for point--like sources with three
 values of the exponent: $\gamma = 1.25$, 1.50 and 1.75.
 For $\gamma = 1.25$ also the cases with sources of linear size $R = 10$~pc and 20~pc are shown.
 \label{fig:ratio_horizon}}
\end{figure}

This is shown in Fig.~\ref{fig:ratio_horizon} that plots the 
ratio $N_A/N_B$ as a function of the horizon
$D_0 = D_H(L_0, \delta = \lambda)$
or $D_* = D_H(L_*, \delta = \lambda)$ assuming indentical sources
or a power--law luminosity distribution. In the second case the curve
is calculated for different
values of the exponent $\gamma$ and assuming point like
or extended sources.
In all cases, increasing the horizon the ratio $N_A/N_B$ grows gradually 
from a value that is approximately unity for small horizon 
($D_H \lesssim 1$~kpc) to a maximum asymptotic value.
For the simple model where all sources are identical,
the asymptotic value of the ratio is simply the
the ratio of the numbers of sources in the volumes of the Milky Way 
subtended by the solid angles $\Delta\Omega_A$ and $\Delta \Omega_B$:
\begin{equation}
 \left (\frac{N_A}{N_B} \right )_{\rm asymp.}^{\rm ident.sources} =
 \left (\int_{\Delta \Omega_A} d^3 x ~f(\vec{x}) \right ) \times 
 \left (\int_{\Delta \Omega_B} d^3 x ~f(\vec{x}) \right )^{-1}
\end{equation}
that is determined by the space distribution of the sources.

\begin{figure}
\begin{center}
\includegraphics[width=12.9cm]{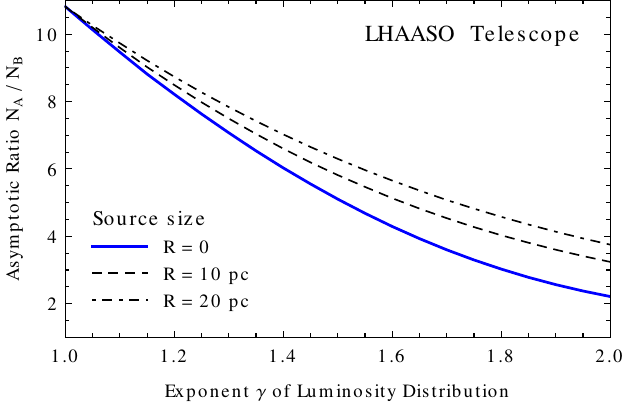}
\end{center}
\caption {\footnotesize
 The ratio $N_A/N_B$ between the number of sources observed
 in region~A and region~B of an EAS telescope takes a
 asymptotic value when the sources have a very high absolute
 luminosity. This asymptotic ratio depends on the space, luminosity
 and size distribution of the sources.
 The figure shows the asymptotic ratio calculated for
 the cylindrically symmetric space distribution discussed in section~\ref{sec:cylindrical},
 sources of identical linear size $R$, and a luminosity
 distribution with the power--law  with exponential cutoff form of Eq.(\ref{eq:lum_dist}),
 The asymptotic ratio is plotted  as a function of the power--law
 slope $\gamma$, and three values of the source size ($R =0$, 10 and 20~pc).
 For a larger exponent $\gamma$  the number of faint sources increases,
 and  the asymptotic ratio decreases.
 \label{fig:asymptotic_ratio}}
\end{figure}

If the luminosity distribution has the power--law
form of Eq.~(\ref{eq:lum_dist}), as already discussed,
the asymptotic ratio depends in the exponent $\gamma$, and
becomes smaller for larger $\gamma$ (approaching unity for $\gamma \to 2$).
This is because the sources that can be resolved at
a distance $D$ from the Earth are those with luminosity larger
than the minimum value $L_{\rm min} (D) \propto \Phi_{\rm min} \, D^2$,
and integrating over the luminosity distribution,
that for $L_{\rm min} \ll L_*$ is a power law $\propto (L/L_*)^{-\gamma}$,
one finds that the contribution of sources at distance $D$ is
\begin{equation}
\frac{dN_{\rm resolved}}{dD} \propto L_*^{(\gamma-1)} \; D^{-2(\gamma-1)} 
\end{equation}
so that the contribution of short distances is enhanced.
The asymptotic ratio $N_A/N_B$ can then be calculated as:
\begin{equation}
 \left (\frac{N_A}{N_B} \right )_{\rm asymp.}^{\rm power-law} =
 \left (\int_{\Delta \Omega_A} d^3 x ~f(\vec{x}) ~|\vec{x} - \vec{x}_\odot|^{-2(\gamma-1)} \right ) \times 
 \left (\int_{\Delta \Omega_B} d^3 x ~f(\vec{x})~|\vec{x} - \vec{x}_\odot|^{-2(\gamma-1)} \right )^{-1} ~.
\end{equation}
The asymptotic ratio decreases for larger values of $\gamma$
because the relative importance of sources at large distances is
suppressed.
This effects is shown in Fig.~\ref{fig:asymptotic_ratio} that plots the asymptotic ratio
$N_A/N_B$ (for the LHAASO telescope) as a function of the exponent $\gamma$.
The two curves in the figure are for point--like objects and for 
sources with a linear size of 20~pc.
For the same luminosity distribution, the ratio $N_A/N_B$ becomes
larger when the sources are more extended in space.
This is a simple consequence of the fact that sources of large angular extension
require a larger flux to be resolved.

The effect we have described is important, because the measured ratio of
the number of resolved sources in different parts of the Galaxy
can be compared with  predictions based on the power--law assumptions for
the luminosity distributions, obtaining an upper limit on the exponent
$\gamma$.

\section{Interpretation of the HAWC and LHAASO gamma--ray observations}
\label{sec:data_analysis}
The main results of our study are contained in four tables
(Tables~\ref{tab:fit_hawc}--\ref{tab:fit_km2a-hig})
that list properties of the Galactic gamma--ray sources estimated
from the data in the HAWC, LHAASO--WCDA, LHAASO--KM2A and
LHAASO--KM2A-high catalogs, using the
method outlined in section~\ref{sec:method}.
These catalogs list sources observed in the energy intervals
[1,10]~TeV (for HAWC and LHAASO--WCDA), [25,100]~TeV (for LHAASO--KM2A)
and [10$^2$,10$^3$]~TeV (for LHAASO--KM2A-high), and the estimated
luminosities are integrated in the same energy intervals. 

The relation between the luminosity of a source and its flux
also depends on the spectral shape. In this study we 
made the simplifying assumption the all sources have the same spectral shape
that is a approximated by a power--law of slope $\alpha = 2.5$ in the [1,10]~TeV energy interval,
and $\alpha = 3.25$ in the [25,100]~TeV and and [10$^2$,10$^3$]~TeV intervals.
For a discussion of the motivation  (and the limitations) of these
choices see our companion paper \cite{lv_spectral_shapes}.

The results in the tables are shown for four models of
the Galactic gamma--ray sources.
Two of the models assume identical point--like sources,
but use the two forms for their space distribution described in
Appendix~\ref{sec:space_distributions}. In the other two models
the luminosity distribution of the sources has the power--law form
of Eq.~(\ref{eq:lum_dist}) with exponent $\gamma=1.25$.
In one case the sources are point--like, in the other they have a linear
size of 20~pc. The choice of $\gamma=1.25$ is motivated by the fact
that larger exponents are disfavored, because they predict
too small $N_A/N_B$ ratios (see Fig.~\ref{fig:asymptotic_ratio}).

The first step of our study is to interpret the number of resolved
sources in the two sky regions~A and~B in order to obtain estimates of
$L_0$ (or $L_*$) and $L_{\rm tot}$
for the hypothesis of identical sources
or a power--law luminosity distribution.
Our analysis method is essentially geometrical, and therefore
in the tables we also give the horizon for the
resolution of a point--like source of luminosity $L_0$ (or $L_*$)
with celestial declination $\delta =\lambda$,
with $\lambda$ the geographic latitude of the telescope.
Points in the sky with this declination culminate at the zenith,
and for them the sensitivity is best, and the horizon is largest.
The horizon for other directions can be be obtained using
the curves in Fig.~\ref{fig:sens_declination} and Eq.~(\ref{eq:thorizon}).

For the hypothesis of identical sources,
the estimated value of the horizon $D_H (L_0)$ is of the order of $\approx 10$~kpc
for all catalogs, indicating that the sources can be resolved in a large,
but limited fraction of the Milky Way volume.
The result that the horizon is approximately the same
for sources observed in different energy intervals
can be understood by noting that the measured $N_A/N_B$ ratios
are  similar in all catalogs. This implies that 
for higher energy intervals the
luminosity of the sources  and the minimum flux for source resolution
decrease at the same rate.

For the hypothesis of a power--law distribution, the
estimated horizon $D_H(L_*)$ is larger (on the order of 20--40~kpc),
indicating that sources wih luminosities $L \gtrsim L_*$
can be resolved in most of the Galactic volume, however the population
of Galactic sources contains many fainter objects that can only be
resolved at shorter distances, reducing the measurable
$N_A/N_B$ ratio.
When the sources have a large linear size, the number of resolved nearby objects
is reduced because they have a larger angular extension
and require a higher flux. Accordingly the $N_A/N_B$ ratio is
obtained with a smaller luminosity $L_*$ and a shorter horizon $D_H (L_*)$.

\begin{figure}
\begin{center}
\includegraphics[width=14.0cm]{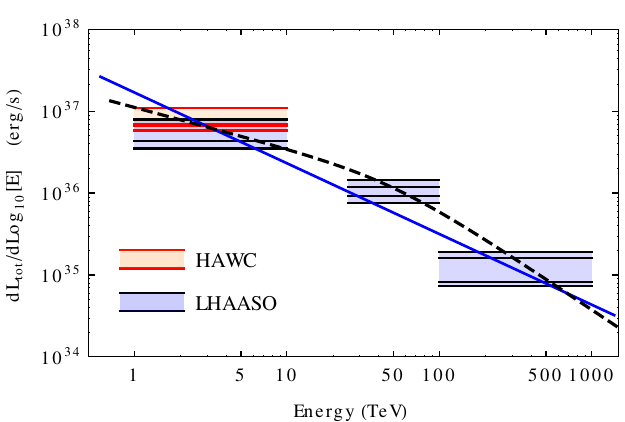}
\end{center}
\caption {\footnotesize
 Total luminosity of the Galactic gamma--ray sources estimated
 from the sky distribution of resolved objects.
 The lines show the estimates obtained using the models
 of the source distributions also used to compute the results
 shown in Tables~\ref{tab:fit_hawc}--\ref{tab:fit_km2a-hig}.
 The lines are not fits, but simple descriptions of the results.
 The solid line is a power--law of form $dL_{\rm tot} /d\log E \propto E^{-0.86}$).
 The dashed line is a broken power--law data with slopes 0.5 and 1.25 below
 and above a break energy of 50~TeV.
 \label{fig:lumtot_fit}}
\end{figure}

Fig.\ref{fig:lumtot_fit}
summarizes all the results on the
estimate of the total luminosity of the Galactic sources $L_{\rm tot}$,
showing them in the form $dL_{\rm ltot}/d\log_{10} E$.
The figure shows that the uncertainty on the total luminosity
is rather small, which is the consequence of a cancellation effect,
because the total Galactic luminosity is the product 
of the source density and the average luminosity of the sources,
and the number of observed resolved sources can be reproduced
by a large density of faint sources visible in a small volume,
or by a smaller density of brighter sources that can be resolved in a larger volume. 

In fig.\ref{fig:lumtot_fit} one of the two curves is a simple power--law of form:
\begin{equation}
 \frac{dL_{\rm tot}}{d\log_{10} E} \approx 1.7 \times 10^{37} ~\left (
 \frac{E}{\rm TeV} \right )^{-0.86} ~\frac{\rm erg}{\rm s} ~,
\end{equation}
and the second is a smoothly broken power--law with
slopes of 0.50 and 1.25 below and above a break of width $w =0.60$:
\begin{equation}
 \frac{dL_{\rm tot}}{d\log_{10} E} \approx 1.1 \times 10^{37}
 ~\left ( \frac{E}{\rm TeV} \right )^{-0.50}
 ~\left [ 1 + \left ( \frac{E}{50~{\rm TeV}} \right )^{1/w}
 \right ]^{-0.25\, w}
 ~\frac{\rm erg}{\rm s } ~.
\end{equation}

The pair of parameters $\{L_0,L_{\rm tot}\}$ or $\{L_*,L_{\rm tot}\}$
determine the luminosity distribution, and therefore it becomes possible
to compute, for any region of the sky, 
the expected number of resolved sources,
and also the total resolved and unresolved fluxes due to the sources.
It is particularly interesting to obtain these estimates for
those regions of the sky where measurements of the diffuse gamma--ray flux have been
obtained, making it possible to compare the estimates with the observations. 

Tables~\ref{tab:fit_hawc}--\ref{tab:fit_km2a-hig} report
for each catalog, in all relevant regions of the sky:
the calculated number of resolved sources,
the fluxes of resolved and unresolved sources, and the ratio
of the latter to the total source flux.
The calculated unresolved source flux can then be compared with
the measured diffuse flux, to estimate the fraction
of the measured diffuse flux that should be attributed
to the contribution of unresolved sources.

In fact, we estimated this unresolved sources contribution
in two different ways.
The first one is simply to compute the ratio
$\Phi_{\rm unresolved}^{\rm model}/\Phi_{\rm diffuse}^{\rm data}$.
In this way, the absolute value of the calculated unresolved flux is used.
A second method is to estimate the unresolved flux from the
sum of the fluxes of all resolved sources in the considered sky region
($\Phi_{\rm resolved}^{\rm data}$) multiplying by the calculated ratio
$\Phi_{\rm unresolved}^{\rm model}/\Phi_{\rm resolved}^{\rm model}$.
This method has the advantage that the effects of an imprecise
description of the detector sensitivity (which determines the absolute value
of the calculated fluxes) to a good approximation cancel out.
The two methods are in reasonably good agreement with each other.

It should be noted that the estimates obtained with the models
are calculated with numerical integrations that implicitly assume
that the sources are smoothly distributed in space,
neglecting the fact that they are discrete. 
The results of the calculation
should therefore be considered as averages 
over all possible configurations of the Galaxy
associated with the same underlying model. 
The fluctuations are to a good approximation
Poissonian for the number of resolved sources,
but can be much larger for the estimate
of the resolved source flux, since the flux of an individual object
is very sensitive to its distance from the Solar system
(diverging at short distances $\propto D^{-2}$).

For the lowest energy interval ([1,10]~TeV), the contribution
of unresolved sources to the diffuse flux in the region
where HAWC has measured the diffuse flux
can reach large values ($\gtrsim 50$\%) only
for a power--law luminosity distribution and for sources of large
linear size (20~pc), while in the other cases it remains significant
but subdominant, accounting for a fraction of 15--30\%.

The LHAASO catalog contains a list of sources identified in the
[1,10]~TeV energy range, and the data can be analysed
with the methods discuss above to estimate the parameters of the source luminosity
distribution $\{L_0, L_{\rm tot}\}$ or $\{L_*, L_{\rm tot}\}$, with results
that are in good agreement with those obtained from the observations of HAWC
(see tables~\ref{tab:fit_hawc} and~\ref{tab:fit_wcda}.
Table~\ref{tab:fit_wcda} gives estimates of the unresolved source flux
in the Inner and Outer--Galaxy sky regions,
however the LHAASO collaboration has not yet released measurements of the
diffuse flux in this energy range. 

At higher energies (in the [25,100]~TeV and [10$^2$,10$^3$]~TeV intervals),
the results of the models can be compared with the
LHAASO measurements of the diffuse flux in the Inner and Outer--Galaxy
sky regions. It turns out that in the Outer--Galaxy region
the contribution of unresolved sources to the diffuse flux
is most likely subdominant. In the [25,100]~TeV interval,
this fraction is in most cases of the order of 10--30\% of the observed flux,
and can approach 50\% or more only if the sources have large size.

The contribution of unresolved sources becomes significantly smaller
in the higher energy interval [10$^2$,10$^3$]~TeV,
reflecting the fact that the resolved
gamma--ray sources have a very soft spectral shape in this energy range.

For the Inner--Galaxy, in the [25,100]~TeV energy range,
the contribution of unresolved sources
could account for a very large fraction ($\gtrsim 50$\%)
or even most of the observed diffuse flux. It should be noted, however, 
that the unresolved source fluxes calculated in this work
refer to the entire sky region considered,
without taking into account the masking adopted in the LHAASO measurement
of the diffuse flux. 
If we tentatively assume that the diffuse flux at high energies
has the same angular dependence as measured by Fermi--LAT
at $E \simeq 10$~GeV, the correction factor
(see Table~\ref{tab:diffuse_angular}) is of the order of 1/0.622,
and the contribution of unresolved sources is reduced accordingly.

As in the case of the Outer--Galaxy, the contribution of unresolved sources
to the diffuse flux is reduced in the highest energy interval
[10$^2$,10$^3$]~TeV, reflecting the softness of the spectra of the observed
sources, and in most models accounts for a fraction of the order of $\sim 20$--30\%
of the observed diffuse flux.

\begin{figure}
\begin{center}
 \includegraphics[width=14cm]{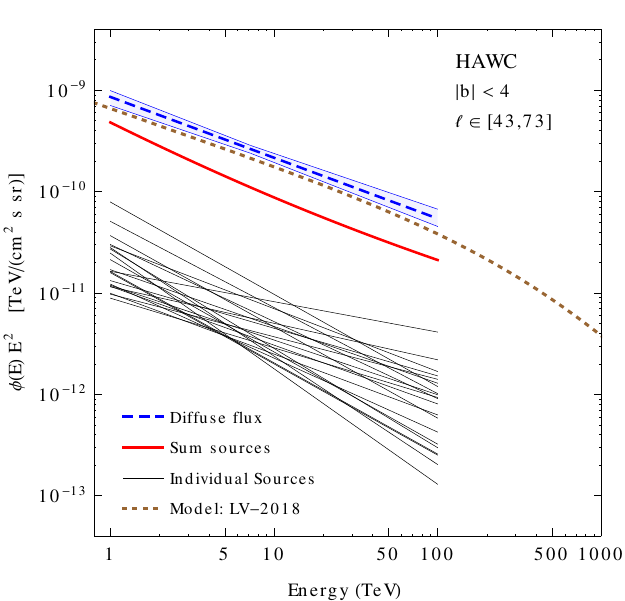}
\end{center}
\caption {\footnotesize
The shaded area shows the best fit and one sigma uncertainty for the
diffuse gamma--ray flux measured by HAWC \cite{HAWC:2023wdq}
in the sky region \{$|b| < 4^\circ$, $43^\circ \le \ell < 73^\circ$.
The thick red line shows the sum of the fits to the 21 sources
in the HAWC third catalog \cite{HAWC:2020hrt} that are in the same sky region,
while the thin lines show the fits to the individual sources.
In the plot the fits to the sources (and their sum)
are divided by the solid angle $\Delta \Omega$ of the region to obtain an average flux.
The thick dotted  line is the prediction of the factorized model
in LV--2018  \cite{Lipari:2018gzn}.
\label{fig:hawc_diffuse_sources}}
\end{figure}

\begin{figure}
\begin{center}
\includegraphics[width=14cm]{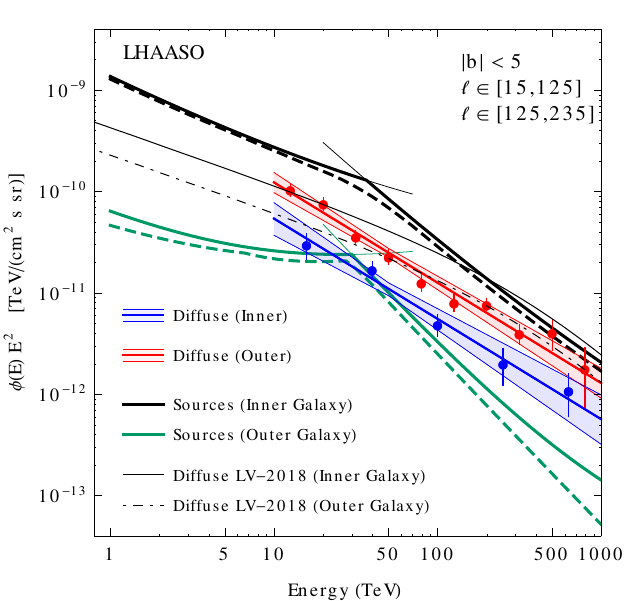}
\end{center}
\caption {\footnotesize
The shaded areas shows the measurements of the diffuse gamma--ray flux
obtained by LHAASO \cite{LHAASO:2023gne} in the sky regions
 \{$|b| < 5^\circ$, $15^\circ \le \ell \le 125^\circ$\} (Inner--Galaxy), and
 \{$|b| < 5^\circ$, $125^\circ \le \ell \le 235^\circ$\} (Outer--Galaxy).
 The thick lines show the sum of the fits of all sources listed 
 in the 1st LHAASO catalog \cite{LHAASO:2023rpg} in the two regions.
 The thick dashed lines show the sum of the combination of fits
 to the sources  observed by both the WCDA and KM2A arrays. 
 The thin lines show the predictions of the factorized model
 in LV-2018 \cite{Lipari:2018gzn} for the two regions.
 \label{fig:lhaaso_diffuse_sources1}}
\end{figure}

A visualization of the relationship between the diffuse flux and the
flux of the resolved sources for the HAWC and LHAASO catalogs
is shown in figures~\ref{fig:hawc_diffuse_sources}
and~\ref{fig:lhaaso_diffuse_sources1}.
Fig.~\ref{fig:hawc_diffuse_sources} shows 
the diffuse flux measured by HAWC in the sky region \{$|b| < 4^\circ$, $43^\circ \le \ell \le 73^\circ$\},
together with the fits to all 21 sources resolved in the same region and the
sum of all these fits. Integrating in the 1--10~TeV energy range, the observed diffuse flux
is $\Phi_{\rm diff} \simeq (3.82^{+0.55}_{-1.10}) \times 10^{-11}$ ~[cm$^{-2}$s$^{-1}$],
while the flux of the resolved sources
$\Phi^{\rm resolved}_{\rm sources} \simeq 1.94 \times 10^{-11}$ ~[cm$^{-2}$s$^{-1}$],
is about half as large. The study of the distribution of the sources
resolved by HAWC suggests [see table~\ref{tab:fit_hawc}] that most (50--80\%)
of the source flux in the region under study is due to contributions from 
resolved objects, so one can conclude that
unresolved sources account for less than half of the measured diffuse flux.
Fig.~\ref{fig:hawc_diffuse_sources} also shows the prediction
of the interstellar emission flux of the factorized model in LV--2018 \cite{Lipari:2018gzn}
which is in reasonably good agreement with the observed flux.
Note that in \cite{HAWC:2023wdq} the HAWC Collaboration compares the diffuse flux data
with the DRAGON base model, which is smaller than the data by a factor of two.

Fig.~\ref{fig:lhaaso_diffuse_sources1} shows the diffuse fluxes measured by LHAASO 
in the Inner and Outer--Galaxy regions, together with the fits to the sources
resolved in these regions, and the sums of all the fits.
In the Inner--Galaxy region 49 sources are measured by WCDA and 55 by KM2A,
with 41 sources measured by both detector arrays, while in the Outer--Galaxy region
9 sources are measured by WCDA and 13 by KM2A with 7 measured by both arrays.

Comparison of these results shows that in the Outer--Galaxy the cumulative
flux of the sources is approximately equal to the measured diffuse flux
at $E \approx 30$~TeV, and then decreases more rapidly with energy.
The estimates shown in Tables~\ref{tab:fit_km2a} and~\ref{tab:fit_km2a-hig} suggest than
the flux of unresolved sources is only a fraction of the combined flux of the resolved sources,
and therefore most of the observed diffuse flux in this region
is generated by interstellar emission,
with the fraction of unresolved sources decreasing with energy because of their
softer spectrum.

In the Inner--Galaxy region the combined flux of all resolved sources is larger than
the measured diffuse flux. Note, however, that the diffuse flux is measured
in only a fraction of the selected region due to masking of known sources.
As discussed above, the solid angle after 
the masking has a larger $\langle |b| \rangle$ than the entire region,
and therefore has also a smaller average flux.
Taking into account this effect
(a correction estimated of the order of 60\% in Table~\ref{tab:diffuse_angular}),
the diffuse flux and the resolved source flux are
approximately equal in the [25,100]~TeV energy range, and the diffuse flux
becomes larger at higher energies.
The results shown in Tables~\ref{tab:fit_km2a} and~\ref{tab:fit_km2a-hig} 
then suggest that unresolved sources contribute a large fraction of
the diffuse flux below 100~TeV, but become subdominant at higher energies.

\section{Galactic Neutrino flux}
\label{sec:neutrinos}

It is well known that the emissions of gamma--rays and high energy neutrinos are
intimately related, and that the simultaneous study of
the fluxes of photons and neutrinos is of fundamental importance to understand
the emission mechanisms and the properties of their sources.

Gamma--rays can be produced by the leptonic and hadronic mechanisms.
In the first case the emission is generated by the radiation
of relativistic electrons and positrons (via Compton scattering on radiation fields
and/or Bremsstrahlung in interactions with ordinary matter)
and in this case there is no corresponding neutrino emission.
In the hadronic mechanism the gamma--rays are emitted
in the decay of mesons created in the inelastic interactions of relativistic protons and
nuclei, with the dominant source being the decay of neutral pions ($\pi^0 \to \gamma \gamma$).
For the hadronic mechanism the gamma--ray emission is accompanied by the
emission of neutrinos generated in the Weak decays of mesons
created in the same hadronic collisions. 
The main neutrino source is the chain decay of charged pions
($\pi^+ \to \mu^+ \nu_\mu$ followed by $\mu^+ \to e^+ \nu_e \overline{\nu}_\mu$ and
charge conjugate modes).
The neutrino flavors are modified during propagation due to oscillations,
so that, assuming that charged pion decay is the primary production  mechanism,
the neutrinos that reach the Earth are to a first approximation a
combination with equal weights of the three flavors 
(see \cite{Lipari:2007su} for more discussion).

The production of charged and neutral pions
is related by isospin symmetry and the pions decay spectra are well known,
therefore it is possible to find a simple relation
connecting the gamma--rays and neutrino emissions.
The spectra of the final state particles generated in the decay of
ultrarelativistic parents have a scaling form and depend
only on the ratio of the energies of the decay product and the parent particle.
The exact forms of the spectra of particles generated in pion decay are known, 
but for the discussion here it is an acceptable approximation to use the simple expressions:
\begin{equation}
\left . \frac{dN}{dx}\right |_{\pi^0 \to \gamma} = 2 ~\delta\left [ x - \frac{1}{2} \right ]
\end{equation}
(with $x = E_\gamma/E_\pi$) for gamma--rays,
and summing over all $\nu$ and $\overline{\nu}$ flavors:
\begin{equation}
\left . \frac{dN}{dx}\right |_{\pi^\pm \to \nu} = 3 ~\delta\left [ x - \frac{1}{4} \right ]~
\end{equation}
(with $x = E_\nu/E_\pi$) for neutrinos.
Assuming that the spectra of the parent charged and neutral pions have the same shape
and relative normalizations $R_\pm$ and $R_0 = 1-R_\pm$,
it is then elementary to derive a simple relation between
the neutrino and gamma--ray emissions.
Neglecting absorption effects, 
the gamma--ray and neutrino fluxes are then related by:
\begin{equation}
\begin{array}{l}
\phi_\gamma (E_\gamma) = \phi_\gamma^{\rm lept} (E_\gamma) + \phi_\gamma^{\rm had} (E_\gamma)
\\[0.35 cm]
\phi_\nu(E_\nu) \simeq 
(R_\pm/R_0) \; 3 \; \phi_\gamma^{\rm had} \left ( 2 \, E_\nu \right ) \simeq
6 \,\phi_\gamma^{\rm had} \left ( 2 \, E_\nu \right )
\label{eq:nu_gamma}
\end{array}
\end{equation}
where in the last equality we have assumed the ratio $R_\pm/R_0 = 2$ predicted by
isospin symmetry.
This equation states the well known fact that if the gamma--ray flux is
generated by the hadronic mechanism, it is accompanied by a neutrino flux
of similar magnitude, because in the second line of Eq.(\ref{eq:nu_gamma}) the factor
of six compensates for the fact that neutrinos are produced at an energy lower
by a factor of two, with spectra that fall off rapidly with energy.
Conversely, a neutrino flux (generated by the standard mechanism described above)
is always accompanied by a gamma--ray flux of comparable magnitude.

In the presence of absorption, the observable fluxes of gamma--rays
are reduced, however absorption during propagation is small or negligible
for energy below 1~PeV \cite{Vernetto:2016alq},
and absorption inside the sources
is also expected to be small in current models of Galactic objects.
By comparing the gamma--rays and neutrino fluxes, it is therefore
in principle possible to determine the relative importance
of the hadronic and leptonic emission mechanisms.

The Galactic neutrino flux observed by IceCube
is the sum of a diffuse component due to interstellar emission
and a second component due to the sum of all neutrino sources.
The only difference with respect to the gamma--ray case
[see Eq.~(\ref{eq:flux_diffuse_composition})] is that
until now no Galactic neutrino sources have yet been identified.

It is expected that at high energies the interstellar
emission is dominated by the hadronic mechanism,
and therefore produces  approximately equal fluxes of neutrinos and gamma--rays.
The nature of the mechanisms that operate in the high energy  sources
is an object of debate.  In  some  classes of sources,
such as Pulsar Wind Nebulae (PWN), the main emission mechanism
is very likely leptonic, but it is possible (and in some sense also inevitable)
that the Galaxy contains sources that generate gamma--rays and neutrinos
via the hadronic mechanism.
This important question  can be addressed comparing the
gamma--ray and neutrino fluxes.

\subsection{The IceCube evidence for Galactic neutrino emission}
\label{sec:nu_icecube}
Recently the IceCube collaboration published \cite{IceCube:2023ame}
evidence  for the emission of high energy neutrinos from the Galactic plane
in the approximate energy range from 1 to 100~TeV.
This result was obtained by using machine learning techniques
and comparing, over the entire celestial sphere,
the IceCube data with three different templates for the diffuse
neutrino flux and with a background--only hypothesis,
leaving  the absolute normalization of the templates  as a free parameter.
This study lead to  evidence for the existence  of a Galactic neutrino flux
at the level of 4.71, 4.37 and 3.96 $\sigma$'s
for the three  templates \cite{IceCube:2023ame}.

One of the three templates (the $\pi^0$ model) is a phenomenological one,
based on the angular distribution of the diffuse gamma--ray
flux measured by the Fermi--LAT (for $E \lesssim 1$~TeV) \cite{Fermi-LAT:2012edv}
and extrapolated to higher energies with a simple 
$\propto E^{-2.7}$ spectrum, assuming factorization of the energy and angular
dependencies of the flux. The other two models (KRA$_\gamma^{5}$ and KRA$_\gamma^{50}$)
are based on predictions of the diffuse gamma--ray flux constructed using 
the cosmic ray transport software package DRAGON including 
a radially dependent diffusion coefficient for cosmic--ray propagation.
This results in a harder spectrum for particles near the center of the Galaxy
($\propto E^{-2.5}$ in the energy range of the Icecube observations),
and in an angular distribution becomes progressively more concentrated
toward the Galactic Center with increasing energy.
The superscripts in the template names  indicate the cutoff energies
(5 and 50~PeV) of the cosmic ray proton spectrum at Earth.

\begin{figure}
\begin{center}
\includegraphics[width=9cm]{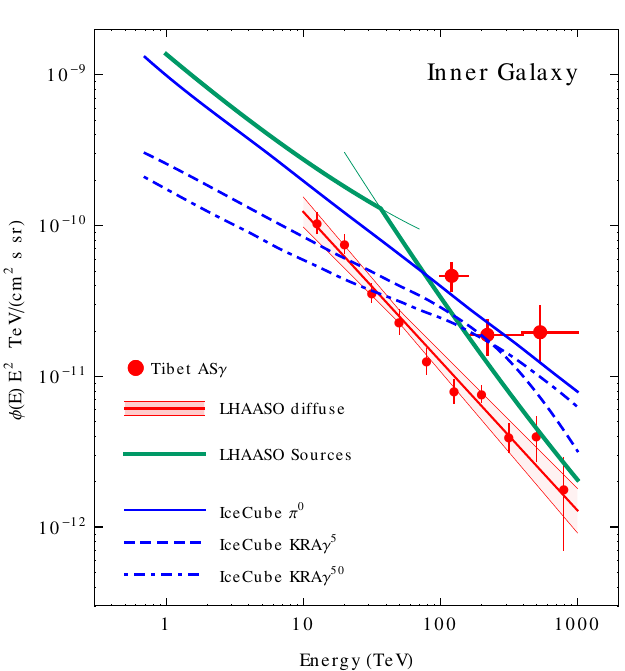}
\vspace{0.25 cm}
\includegraphics[width=9cm]{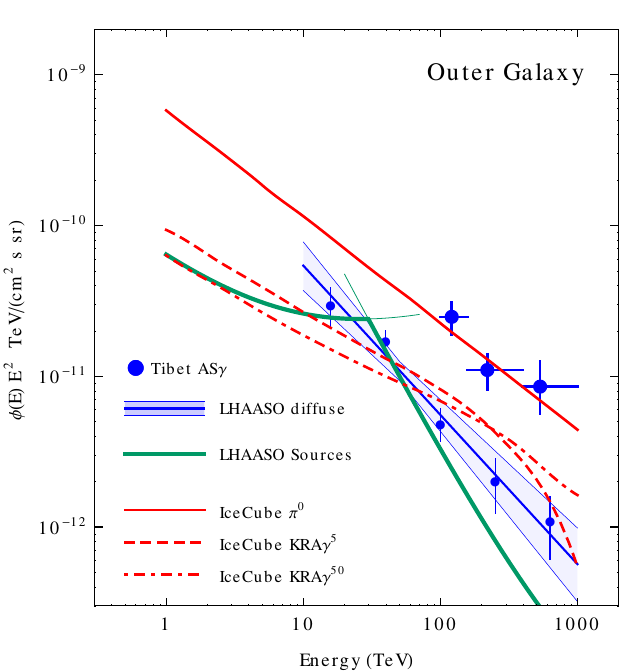}
\end{center}
\caption {\footnotesize
Comparison of the average Galactic gamma--ray and neutrino fluxes
obtained by Tibet--AS$\gamma$  \cite{TibetASgamma:2021tpz},
LHAASO \cite{LHAASO:2023rpg,LHAASO:2023gne} and IceCube \cite{IceCube:2023ame}.
The IceCube results are shown in the form of gamma--ray fluxes
calculated  with  Eq.(\ref{eq:nu_gamma}) and 
using as input the  neutrino fluxes 
shown in  Fig.S8  of \cite{IceCube:2023ame}.
These neutrino  fluxes are  the three templates
averaged over the Tibet--AS$\gamma$
Inner and Outer Galaxy regions with the absolute normalization obtained from  best fits
to the data  over the entire celestial sphere.
The top (bottom)  panel  shows the IceCube results for the Inner--Galaxy (Outer--Galaxy)
region,  together with the  measurements of the  Tibet--AS$\gamma$ and LHAASO
telescopes.
Note that the  definitions of the sky regions differ for the two gamma--ray telescopes.
see Table~\ref{tab:diffuse_angular}).
\label{fig:icecube_inner}}
\end{figure}

It is not easy to compare the IceCube neutrino results
(based on all sky  templates of the neutrino flux) 
with those of the gamma--ray telescopes
(that measure the flux in  some limited regions of the sky).
However, the supplemental material of the IceCube paper \cite{IceCube:2023ame}
shows (in Fig.S8) the spectra of the three templates (with the best fit normalization)
averaged over the Tibet--AS$\gamma$ Inner and Outer--Galaxy regions.
These spectra, transformed into gamma--ray spectra using Eq.(\ref{eq:nu_gamma}) 
are shown in Fig.~\ref{fig:icecube_inner} together with
the Tibet--AS$\gamma$ and LHAASO telescopes measurements.
Examining this figure, several comments are in order:

\vspace{0.25 cm}
\noindent [A]
The average fluxes measured by Tibet--AS$\gamma$ \cite{TibetASgamma:2021tpz}
and LHAASO \cite{LHAASO:2023rpg,LHAASO:2023gne}
in their respective Inner--Galaxy regions are expected
to be very similar to each other, because the LHAASO region is
larger  (the two regions have the same  latitude range $|b| \le 5^\circ$, 
but longitude ranges $15^\circ \le \ell \le 125^\circ$ for LHAASO versus
$25^\circ \le \ell \le 100^\circ$ for Tibet--AS$\gamma$) but the extension
is  partly toward the Galactic Center and partly toward the Galactic
anti--center.
The total gamma--ray flux observed by LHAASO in this sky region is
dominated by the contribution of photons emitted by identified sources.
Summing the resolved source contribution with the measured diffuse flux,
and correcting for the effect of the masking [see Table~\ref{tab:diffuse_angular}]
the LHAASO total gamma--ray flux can be reconciled 
with the flux  measured by Tibet--AS$\gamma$.
Note however that this conclusion requires the assumption that, 
contrary to what is discussed in  \cite{TibetASgamma:2021tpz},
most of the  gamma--ray flux observed by  the Tibet--AS$\gamma$ telescope 
is  due to the contribution of unresolved sources.

\vspace{0.25 cm}
\noindent [B]
The Outer--Galaxy regions for Tibet--AS$\gamma$ and LHAASO are 
quite different, because the one for Tibet--AS$\gamma$ extends
to smaller longitudes ($50^\circ \le \ell \le 200^\circ$
versus $125^\circ \le \ell \le 235^\circ$ for LHAASO.
For LHAASO, according to the analysis presented in
the previous section,
the total gamma--ray flux in the Outer--Galaxy region 
consists of approximately equal contributions from sources and interstellar emission,
with the latter dominating the flux observed as diffuse.
In the flux measured by Tibet--AS$\gamma$ the contribution of sources
is estimated to be larger than for LHAASO,
with the  interstellar emission accounting for a fraction of the  order of 25\% of the total flux.

\vspace{0.25 cm}
\noindent [C] The crucial question that can be addressed by comparing
the gamma--ray and neutrino fluxes, is what fraction of the
gamma--ray emission from the sources
is produced  by the hadronic mechanism, and therefore accompanied by
neutrinos. The high neutrino flux observed 
in the Tibet--AS$\gamma$  Inner--Galaxy region for a neutrino energy of the order of 50~TeV
(gamma--ray energy of the order of 100~TeV) is an intriguing hint
that an important fraction of the source emission could be generated
by the hadronic mechanism, but a robust conclusion requires a more in depth study.

\vspace{0.25 cm}
\noindent [D] The $\pi^0$ and the two KRA$\gamma$ templates have two main
differences,  with the KRA$\gamma$ templates predicting:
(i) much harder spectra in directions toward the Galactic Center,
and (ii) a significantly larger Inner/Outer ratio.
The gamma--ray observations find spectra that are soft
and have a shape independent of direction.
For example, the diffuse flux measurements of LHAASO above 25~TeV were
fitted with power--law spectra of  equal slopes
($2.99\pm 0.08$ and $2.99\pm 0.14$) in the Inner and Outer Galaxy.
In addition, the observed  Inner/Outer  ratios (for both Tibet--AS$\gamma$ and LHAASO
are smaller than the predictions of the KRA$\gamma$ templates.
Accordingly, the hypothesis at the basis of the KRA$\gamma$ models
seems to be disfavored.

\section{Summary and Outlook}
The study of the Galactic diffuse gamma--ray flux is of great importance,
since the flux encodes information about the 
cosmic ray spectra in different regions of the Milky Way,
and the extension of the measurements to very high energies in the TeV and PeV
energy ranges is of crucial importance.
However, in order to interpret the recent observations of high energy telescopes
it is necessary to identify and subtract the contribution of unresolved sources
to what is measured as a diffuse flux.
The measurement of the unresolved source flux is of course also
of great importance in itself and is required for a complete understanding of
the properties of the Galactic high energy accelerators. 

The recent publications of LHAASO \cite{LHAASO:2023gne} and HAWC \cite{HAWC:2023wdq},
which have reported measurements of the diffuse flux in some regions of the
Galactic disk, note in the discussion of their results
that the data appear to show a significant excess
over theoretical expectations that can be explained
as the contribution of unresolved objects.

However, the evidence for the existence of an excess
in the diffuse gamma--ray flux at high energies
is  based on the comparison of the data with some specific theoretical
models, while other models predict larger fluxes.
For example, in this work
we have shown that the predictions of the
LV--2018 model \cite{Lipari:2018gzn}
are in reasonably good agreement with the HAWC and LHAASO observations,
and are in fact somewhat larger than the LHAASO data (see also \cite{Vecchiotti:2024kkz}).
Therefore the size of the contribution of unresolved sources
to the observed diffuse gamma--ray flux remains 
a very important open problem that can be addressed in several ways.

The approach followed in this paper is based on
the study of the angular distributions of the resolved sources. 
These distributions are determined in part by the (direction dependent) sensitivity
of the telescopes but also encode the space and luminosity distributions
of the Galactic gamma--ray sources.
A sufficiently accurate description of these distributions allows 
the fluxes of unresolved sources to be  calculated.
In this paper, to extract this information from the data,
we make two main simplifying assumptions:
(i) that the space
and luminosity distributions of the Galactic gamma--ray sources are factorized,
and (ii) that their space distribution is known,
at least to a reasonably good approximation.
With these assumptions it is possible to
infer the parameters of the luminosity distribution of the Galactic gamma--ray sources
from the measured angular distribution of those that are resolved,
and then to compute the unresolved source flux for any region of interest.
It is important to note that this program does not require any measurement
of the distances of the resolved sources.

Two major difficulties are that the total number
of resolved sources is still relatively small,
and that the detailed form of the sky distribution of the sources
is not known very precisely.
In fact the angular distribution of the resolved sources
is determined by the spiral structure of the Milky Way, and by the shape
of the spiral arms in the vicinity of the Earth. Current models of the spiral structure
cannot describe the details of the longitude and latitude
distributions of the data. On the other hand, the general structure
of the space distribution of the sources, i.e.  a thin disk with a density
decreasing exponentially with galactocentric radius is reasonably well known.
For these reasons, in this paper we have performed a first approximation study,
by dividing the sky into two regions of equal solid angle and sensitivity,
one centered (to a good approximation) on the Galactic anticenter (region~B),
and the other as close as possible to the Galactic center (region~A),
and comparing the number of resolved sources
in the two regions. In this way the statistical errors are reduced, and the
uncertainties in the description of the space distribution of the sources play
a less important role. Of course, this choice has the important limitation
that it allows to determine only two parameters for the luminosity distribution
of the sources, which can be chosen as the total luminosity of
the ensemble of all Galactic sources, and a
characteristic luminosity for the individual sources.

The LHAASO and HAWC catalogs observe 3--4 times more
sources in the sky region~A toward the Galactic center.
Since the density of stars and other astrophysical objects, changes with
the galactocentri radius only on a scale of several kpc,
it is easy to see that the large $N_A/N_B$ ratio implies that gamma--ray sources
can be resolved out to distances of several kpc.
In addition, this result allows to place limits on
the contribution of very faint objects to the unresolved source flux.
This is because faint objects can only be resolved at short distances, and if they were
too numerous, they would push the ratio $N_A/N_B$ down toward unity.

In this work we have used a model of the source luminosity distribution that
is a power--law with an exponential cutoff, and found that the observation
of a large asymmetry in the number of resolved sources in the
regions~A and~B can be used to set an upper limit on the exponent
that controls the low luminosity end of the distribution $\gamma \lesssim 1.25$.

We have used the ratio $N_A/N_B$ in the LHAASO and HAWC catalogs to
estimate the horizon for source resolution, and thus the
luminosity of the individual gamma--ray sources, and then we used the absolute numbers of
resolved sources to estimate the total luminosity of all Milky Way sources.
This total luminosity can be reconstructed with an uncertainty of about a factor
of two, and (in units erg/s) is approximately 
$1.5 \times 10^{37}$, 
$4 \times 10^{36}$ and 
$4 \times 10^{35}$ in the energy intervals [1,10], [10,100], and [10$^2$,10$^3$]~TeV.

These results were then used to estimate the flux of unresolved sources,
and it was found that,
with the  possible exception of directions close to the Galactic Center,
the contribution of unresolved sources is likely to be subdominant in what
is measured as the diffuse gamma--ray flux.
The simplest way to obtain a first estimate of
the unresolved source flux  is to use the measured 
flux of the resolved sources, and to note that if
the horizon for source resolution is of the order of $\approx 10$~kpc,
the flux from sources beyond the horizon is  only a fraction of the
flux from the sources inside the horizon.

More detailed studies of the properties of the Galactic gamma--ray sources are
not only desirable but in fact necessary to verify the results and conclusions
obtained in the simple study developed here.
A more precise description of the spiral structure of the Milky Way,
capable of reproducing the observed features in the latitude and longitude
distributions of the sources is very desirable, and would allow
the shape of the luminosity distribution of the sources to  be
studied in more detail.
In would also be very interesting to perform separately these studies for
different classes of sources, which are also likely to have different
extensions.
It is also clear that the inclusion in future studies of the
available information on the distances for a subset of the
sources would be very valuable.

\vspace{0.35 cm}
\noindent {\bf Acknowledgments} We are grateful to Cao Zhen, Felix Aharonian,
You Zhi Yong, Xi Shao Qian and Zha Min for useful discussions.

\clearpage

\appendix

\section{Space distribution of the Milky Way gamma--ray sources}
\label{sec:space_distributions}

\subsection{Cylindrically symmetric model}
\label{sec:cylindrical}
In this work to describe the space distribution of the Milky Way sources we will use
a simple form symmetric for rotations around the $z$ axis and for
reflections on the ($xy$) Galactic plane:
\begin{equation}
f(\vec{x}) = f_r (r) \, f_z (z)
\end{equation}
For the radial part we used the form introduced by
Yusifov and Kucuk \cite{Yusifov:2004fr} and then used by several other authors:
\begin{equation}
 f_r(r) = K_r ~\left ( \frac{r + r_0}{r_\odot + r_0} \right )^a ~\exp\left [-\beta \left (
 \frac{r-r_\odot}{r_\odot + r_0} \right ) \right ]
\label{eq:f_radial}
\end{equation}
This form depends on three parameters ($r_0$, $a$ and $\beta$) with
$K_r$ is a normalization factor that can be calculated numerically.

For the $z$ dependence we used an exponential form, with a correction to avoid
a discontinuity in the derivative at $z =0$:
\begin{equation}
f_z(z) = K_\zeta ~e^{z/Z} \left ( 1 + e^{2 z /(Z \, \zeta)} \right )^{-\zeta}
\label{eq:fz}
\end{equation}
with $K_\zeta$ a normalization factor (to insure normalization to unity):
\begin{equation}
 K_\zeta = \frac{1}{Z} \; \frac{4 \Gamma(\zeta)}{\zeta \, \Gamma^2(\zeta/2)}
 \simeq \frac{1}{Z} \;\left ( 1 + \frac{\pi^2 \zeta^2}{24} + \ldots \right )~.
\end{equation}
In the limit of $\zeta \to 0$ one has $f_z(z) \to e^{-|z|/Z}/Z$, and the
function has a cusp at $z = 0$, while for
$\zeta = 1$ one has $f_z(z) = [\pi\, Z \, \cosh(z/Z)]^{-1}$.
In general the second derivative of $f_z (z)$ at $z=0$ is:
\begin{equation}
\left . \frac{d^2 f_z(z)}{dz^2} \right |_{z=0} = - \frac{1}{\zeta \, Z^2} ~.
\end{equation}

For our numerical studies we have tested several
set of values for the parameters,
but the results we are presenting are based on the
parameters suggested by Lorimer et al.\cite{Lorimer:2006qs}
to describe the observations of
the Parkes 20-cm multibeam pulsar survey of the Galactic plane
($r_0 = 0.55$~kpc, $a=1.64$, $\beta = 4.01$) for the radial distribution and 
$Z = 0.18$~kpc for the $z$ distribution.

\subsection{Spiral model of the Milky Way} 
\label{sec:spirals}
Our model for the spiral structure of the Galaxy is based on the analysis
by Reid et al. \cite{Reid:2019} of trigonometric parallax and proper motion
measurements of molecular masers associated with very young high-mass stars.
These objects trace the high-mass star-forming regions of our Galaxy,
where gamma ray sources such as PWN and SNR are more likely to be found.
The maser measurements have been performed in the 1$^{st}$, 2$^{nd}$ and 3$^{rd}$ Galactic quadrants.
Considering also the well-established arm tangencies in the 4$^{th}$
Galactic quadrant, the authors developed a model with four spiral arms and some
extra arm segments and spurs. Each arm $k$ is described by a logarithmic spiral curve:
\begin{equation}
r = R_k \; \exp\left [-(\beta-\beta_k) \tan \psi_k \right ]
\end{equation}
where $r$ is the distance from the Galactic center, $\beta$ the azimuth angle
and $\psi$ the pitch angle. The values of $\psi$, $R_k$ and $\beta_k$ parameters have been
found by fitting the spiral function to the data.
The best fit pitch angles are not constant along the spiral lines,
but change at some azimuth angles. The best-fit parameter values are given in Table 1 of Reid et al.
for a limited range of azimuth angles. Beyond this range we extracted them from
Fig.~1 of the same paper, and we extrapolated the spiral lines outside the region shown
in the figure using the same pitch angles of the last segment of the curve. 
In our model we consider four main arms: Norma--Outer, Scutum--Centaurus, Sagittarius--Carina and
Perseus, plus the local Orion arm, a small spur in which the Sun is located. 

Fitting the maser data, Reid et al. found that the arms width in the Galactic
plane increases with the distance $r$ from the Galactic center and for $r >$ 3~kpc can be described by the expression:
\begin{equation}
w(r) = 0.336 + 0.036 \, (r-r_\odot) 
\label{eq:arm_w}
\end{equation}
where $w(r)$ is in kpc and $R_\odot$ = 8.15~kpc. 

Using this geometric model of the Galaxy, we have "filled" the arms (projected on the Galactic plane)
with a source number consistent with our model with cylindrical simmetry (described in Appendix A1),
normalizing the source number of the two models for each distance $r >$ 4~kpc from the Galactic center. 
More explicitly, if there are $N$ spiral arms at the distance $r$, then each arm
contains (between $r$ and $r+dr$) 1/$N$ of the number of sources located in the ring of
area 2$\pi$ $r dr$ of the cylindrical model. The sources are then distributed on
the Galactic plane, perpendicularly to each spiral line, according to the expression (\ref{eq:arm_w}).

For the Galactic center region ($r < 4$~kpc), we assume a bar structure
with a source density (projected onto the Galactic plane) according to the long-bar model
by Wegg et al. (Eq.9 in \cite{Wegg:2015}):
\begin{equation}
 \rho_b (x_b,y_b) \propto
 \exp \left \{ - \left [ \left ( \frac{x_b}{X_b} \right )^{g} +
 \left ( \frac{y_b}{Y_b} \right)^{g} \right]^\frac{1}{g} \right \} ~
 G \left ( \frac{r_b-R_b}{\sigma_b} \right )
\end{equation}

where $x_b$ and $y_b$ are right-handed galactocentric coordinates with the $x_b$ axis oriented along the major axis of the bar, which is assumed to be inclined 
by 30 degrees with respect to the line connecting the Galactic center to the Sun and $r_b = (x_b^2+y_b^2)^{1/2}$.
$G(x)$ is a function describing the outer cutoff of the bar along the major axis:
$G(x)$ = 1 for $x \le$~1, and $G(x) = \exp(-x^2)$ for $x >$~k1.
We assume no cutoff toward the Galactic center.
According to \cite{Wegg:2015}, the values of the parameters are: $X_b$ = 3.05~kpc, $Y_b$ = 0.68~kpc, $g$ = 2.27, $R_b$ = 3.85~kpc and $\sigma_b$ = 0.72~kpc.

The source density of the bar is normalized by requiring that the total number of sources (spirals plus bar) 
be equal to the number of sources in the cylindrical model. 
The source density is adjusted to smoothly connect the bar to the spiral arms.

For the source distribution along the $z$ coordinate, we use (for spiral arms and bar) 
an exponentially decreasing density as $exp(-|z|/z_0)$ with $z_0=0.18$~kpc, 
consistent with the previously described cylindrical model (except for the cusp correction, which is not applied here).

\begin{figure}
\begin{center}
\includegraphics[width=14cm]{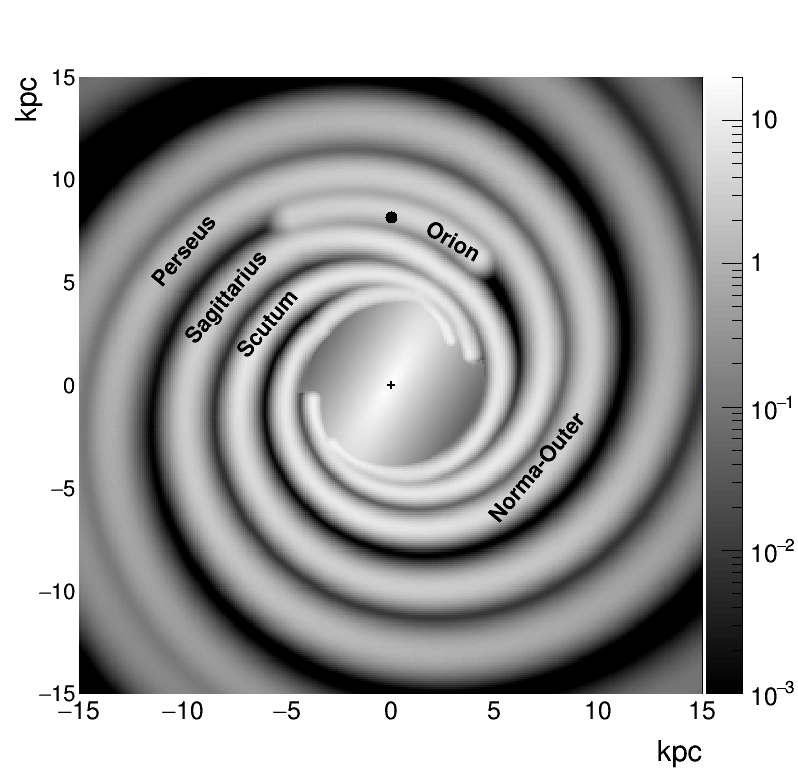}
\end{center}
\caption {\footnotesize
 Model of the spiral structure of the Milky Way described in Appendix~\ref{sec:spirals}.
 The position of the Solar System is indicated by the black dot.
\label{fig:galaxy_map}}
\end{figure}

Fig.~\ref{fig:galaxy_map} shows the source density of the model projected on the Galactic plane.

\clearpage

%%%%%%%%%%%%%%%%%%%%%%%%%%%%%%%%%%%%%%%%%%%%%%%%%%%%%%%%%%%%%%%%%%%%%%%%%%%%
%% Table. HAWC sources 
%%%%%%%%%%%%%%%%%%%%%%%%%%%%%%%%%%%%%%%%%%%%%%%%%%%%%%%%%%%%%%%%%%%%%%%%%%%%

\begin{table}
\caption{\footnotesize
Estimate of gamma--ray sources properties obtained
from the observations of the HAWC telescope \cite{HAWC:2020hrt}.
The top part of the table lists information obtained by the telescope:
the total number of resolved Galactic sources and the number of
resolved sources in the sky regions~A and~B [defined in Eq.~(\ref{eq:acc2})].
More experimental results are given for a smaller sky region
(the Inner--Galaxy region) where the HAWC Collaboration has
obtained a measurement of the average diffuse flux:
the number of resolved sources in the region,
the best fit for the diffuse flux, 
and the total flux from  resolved sources in the region.
The bottom part of the table lists results
obtained interpreting the data in the framework of some models
discussed in this work, defined by the sources space and luminosity
distributions and the source linear size $R$.
The four columns are for four models. In the first two 
the Galactic sources are assumed to be 
point--like and identical, with luminosity $L_0$,
but two different space distributions are used,
one cylindrically symmetric, and one including a description of
the spiral structure of the Milky Way [see Appendix~\ref{sec:space_distributions}].
In the other two models the luminosity distribution of the sources
is a power--law of index $\gamma = 1.25$ with cutoff luminosity
$L_*$ [see Eq.~(\ref{eq:lum_dist})].
The horizon listed in the table
is the maximum distance to resolve a point--like source
with celestial declination $\delta = \lambda$ and luminosity
$L_0$ (for the identical sources model) or $L_*$ (for the power--law
luminosity distribution models).
The last two rows in the table compare the flux of unresolved sources
estimated (with two different methods, see main text) in the model
with the observed diffuse flux.
\label{tab:fit_hawc}}

\vspace{0.40 cm}
\renewcommand{\arraystretch}{1.5}

\begin{tabular}{ | l || c | c | c | c |}
 \hline
\multicolumn{5}{|l|}
{HAWC telescope. Energy Interval [1--10]~TeV} \\
\multicolumn{5}{|l|}
 {All Sky: $N_{\rm sources} = 63$ ~~~~Regions A: $N_{A} = 49$,
 ~~~~Region B: $N_{B} = 14$. } \\
\hline
\multicolumn{5}{|l|}
{Inner Galaxy [($|b| < 4^\circ$), ($43^\circ \le \ell \le 73^\circ)$] } \\
\multicolumn{5}{|l|} {$N_{\rm sources}^{\rm data} = 21$} \\
\multicolumn{5}{|l|} {$\Phi_{\rm resolved}^{\rm data} = 1.94 \times 10^{-11}$ ~[cm$^{-2}$s$^{-1}$].} \\
\multicolumn{5}{|l|} {$\Phi_{\rm diffuse}^{\rm data} = (3.82^{+0.55}_{-1.10}) \times 10^{-11}$ ~[cm$^{-2}$s$^{-1}$]} \\ 
\multicolumn{5}{|l|} {$\Phi_{\rm diffuse}^{\rm data}/\Phi_{\rm resolved}^{\rm data} = 2.0^{+0.3}_{-0.6}$} \\

\hline
\hline
\hline

\multicolumn{5}{|l|} {Modeling of gamma--ray sources} \\
\hline
~~ & Model~1 & Model~2 & Model~3 & Model 4 \\
\hline
Luminosity distribution & Identical sources & Identical sources &
P.L. ($\gamma = 1.25$) & P.L ($\gamma =1.25$) \\
Space distribution & Smooth (PWN) & Spirals & Smooth (PWN) & Smooth (PWN) \\
Source size & $R = 0$ & $R = 0$ & $R = 0$ & $R= 20$~pc \\ 
\hline
\multicolumn{5}{|l|} {Global properties of Galactic gamma--ray sources} \\
\hline
Horizon [kpc] & 
$	6.6	^{+	1.9	}_{	-2.2	}$
&
$	7.7	^{+	2.0	}_{	-2.2	}$
&
$	13.9	^{+	8.7	}_{	-5.8	}$
&
$	9.9	^{+	5.5	}_{	-4.1}$
\\
$L_0$ or $L_*$ [$10^{33}$~erg/s] & 
$	6.4	^{+	4.2	}_{	-3.5	}$
&
$	8.5	^{+	4.1	}_{	-3.2	}$
&
$	24^{+44}_{	-17}$
&
$	14.3	^{+	20.	}_{	-9.3	}$
\\$L_{\rm tot}$ [$10^{36}$~erg/s] & 
$	5.8	^{+	0.31	}_{	-0.25	}$
&
$	7.0	^{+	0.32}_{	-0.26	}$
&
$ 6.6	^{+	1.9	}_{	-0.8	}$
&
$	10.8	^{+	3.9	}_{	-0.5	}$
\\
\hline
\hline
\multicolumn{5}{|l|} {Modeling sources in Inner--Galaxy} \\
\hline
$N_{\rm sources}^{\rm model}$ &
$17^{+2}_{-4}$ & $23^{+2}_{-5}$ & $15\pm 1$ & $17^{+1}_{-3}$
\\
$\Phi_{\rm resolved}^{\rm model}$~[$10^{-11}$\;(cm$^2$s)$^{-1}$]~~ & 
$	2.2	^{+	0.5	}_{	-0.6	}$
&
$	3.3	^{+	0.6	}_{	-0.7	}$
&
$	2.7	^{+	1.3	}_{	-0.9	}$ 
&
$	3.4 ^{+0.7}_{-0.5}$ \\
$\Phi_{\rm unresolved}^{\rm model}$~[$10^{-11}$\;(cm$^2$s)$^{-1}$]~~ & 
$	0.91	^{+	0.49	}_{	-0.32	}$
&
$	0.95	^{+	0.50	}_{	-0.33	}$
&
$	0.91	^{+	0.50	}_{	-0.29	}$
&
$	2.23	^{+	2.53	}_{	-0.98	}$
\\
$f_u = \Phi_{\rm unresolved}^{\rm model}/\Phi_{\rm all}^{\rm model}$ & 
$	0.29	^{+	0.18	}_{	-0.11	}$
&
$	0.22	^{+0.15	}_{-0.12	}$
&
$	0.25	^{+	0.20	}_{-0.12	}$
&
$	0.40	^{+	0.22	}_{	-0.16	}$ \\
\hline
$\Phi_{\rm unresolved}^{\rm model}/\Phi_{\rm diffuse}^{\rm data}$ & 
$0.24^{+0.08}_{-0.12}$ &
$0.15^{+0.06}_{-0.12}$ &
$0.24^{+0.08}_{-0.12}$ &
$0.58^{+0.45}_{-0.33}$ 
\\
$[\Phi_{\rm sources}^{\rm data} \, f_u/(1-f_u)] /\Phi_{\rm diffuse}^{\rm data}$ ~~&
$0.21^{+0.18}_{-0.12}$ &
$0.17^{+0.21}_{-0.09}$ &
$0.17^{+0.19}_{-0.11}$ &
$0.33^{+0.39}_{-0.21}$ \\
\hline
\end{tabular}
\end{table}

%%%%%%%%%%%%%%%%%%%%%%%%%%%%%%%%%%%%%%%%%%%%%%%%%%%%%%%%%%%%%%%%%%%%%%%%%%%%
%% Table. LHAASO-WCDA 
%%%%%%%%%%%%%%%%%%%%%%%%%%%%%%%%%%%%%%%%%%%%%%%%%%%%%%%%%%%%%%%%%%%%%%%%%%%%
 
\begin{table}
\caption{\footnotesize
Estimate of gamma--ray sources properties obtained
from the observations of the LHAASO-WCDA telescope \cite{LHAASO:2023rpg}
in the energy range [1,10]~TeV.
The structure of this table is the same as for
table~\ref{tab:fit_hawc}, with the top part 
listing experimental results, and the bottom part
showing theoretical estimates obtained
using the same models.
Both experimental results and theoretical estimates
are shown for the two sky regions (Inner--Galaxy and Outer--Galaxy regions)
where LHAASO has published measurements of the
diffuse flux in the higher energy range ($E > 25$~TeV).
A measurement of the diffuse flux   in the WCDA energy range  is not
yet available at the moment.
\label{tab:fit_wcda}}

\vspace{0.35cm}
\renewcommand{\arraystretch}{1.4}
\begin{tabular}{ | l || c | c | c | c |}
 \hline
 
\multicolumn{5}{|l|}
{LHAASO--WCDA telescope. Energy Interval [1--10]~TeV} \\
\multicolumn{5}{|l|}
 {All Sky: $N_{\rm sources} = 65$ ~~~~Regions A: $N_{A} = 51$,
 ~~~~Region B: $N_{B} = 14$. } \\

\hline

\multicolumn{5}{|l|}
{Inner Galaxy [($|b| < 5^\circ$), ($15^\circ \le \ell \le 125^\circ)$] } \\
\multicolumn{5}{|l|} {$N_{\rm inner}^{\rm data} = 49$} \\
\multicolumn{5}{|l|} {$\Phi_{\rm resolved}^{\rm data} = 25.8 \times 10^{-11}$ ~(cm$^2$s)$^{-11}$.} \\
\hline
\multicolumn{5}{|l|} {Outer--Galaxy
[($|b| < 5^\circ$), ($125^\circ \le \ell \le 235^\circ)$] } \\
\multicolumn{5}{|l|} {$N_{\rm outer}^{\rm data} = 9$} \\
\multicolumn{5}{|l|} {$\Phi_{\rm resolved}^{\rm data} = 1.38 \times 10^{-11}$ ~(cm$^2$s)$^{-1]}$.} \\

\hline
\hline
\hline

\multicolumn{5}{|l|} {Modeling of gamma--ray sources} \\
\hline
~~ & Model~1 & Model~2 & Model~3 & Model 4 \\
\hline
Luminosity distribution & Identical sources & Identical sources &
P.L. ($\gamma = 1.25$) & P.L ($\gamma =1.25$) \\
Space distribution & Smooth (PWN) & Spirals & Smooth (PWN) & Smooth (PWN) \\
Source size & $R = 0$ & $R = 0$ & $R = 0$ & $R= 20$~pc \\ 

\hline
\multicolumn{5}{|l|} {Global properties of Galactic gamma--ray sources} \\
\hline

Horizon [kpc] & 
$	9.2 \pm 3.2$
&
$	9.0	^{+	3.7	}_{	-2.5	}$
&
$	27.	^{+	49.	}_{	-15.	}$
&
$	20.	^{+	29.	}_{	-11.	}$
\\
$L_0$ or $L_*$ [$10^{33}$~erg/s] & 
$	8.0	^{+	6.5	}_{	-4.5	}$
&
$	7.5	^{+	7.5	}_{	-3.5	}$
&
$	67.^{+90}_{	-30}$
&
$	38.	^{+50.	}_{	-30.	}$
\\
$L_{\rm tot}$ [$10^{36}$~erg/s] & 
$	4.3	^{+	0.9	}_{	-0.6	}$
&
$	3.5	^{+	1.2	}_{	-0.6	}$
&
$	6.7	^{+	11	}_{	-2.6	}$
&
$	7.9	^{+5.4	}_{	-1.0	}$
\\
\hline
\hline
\multicolumn{5}{|l|} {Modeling LHAASO--WCDA sources in Inner--Galaxy } \\
\hline
$N_{\rm inner}^{\rm model}$ &
$44^{+4}_{-8}$ &
$43^{+4}_{-7}$ &
$43^{+4}_{87}$ &
$44^{+4}_{-8}$ \\
$\Phi_{\rm resolved}^{\rm model}$ ~~[$10^{-11}$\;(cm$^2$s)$^{-1}$] & 
$	6.4	^{+	2.3	}_{	-2.1	}$
&
$	7.1	^{+	3.5	}_{	-2.1}$
&
$	11.4	^{+	15.	}_{	-6.1	}$ 
&
$	11.8	^{+	12	}_{	-4.5	}$ \\
$\Phi_{\rm unresolved}^{\rm model} $ ~~[$10^{-11}$\;(cm$^2$s)$^{-1}$] & 
$	2.2	^{+	0.8	}_{	-0.5}$
&
$	1.7	^{+	0.6	}_{	-0.6	}$
&
$	2.1	^{+	0.9	}_{	-0.7	}$
&
$	3.7	^{+	3.0	}_{	-2.0	}$
\\
$f_u = \Phi_{\rm unresolved}^{\rm model}/\Phi_{\rm all}^{\rm model}$ & 
$	0.25	^{+	0.15	}_{	-0.09}$
&
$	0.20	^{+	0.12	}_{	-0.1-	}$
&
$	0.16	^{+	0.20	}_{	-0.12	}$
&
$	0.24	^{+	0.20	}_{	-0.16	}$
\\
\hline
\hline
 \multicolumn{5}{|l|} {Modeling LHAASO--WCDA sources in Outer--Galaxy. } \\
\hline
$N_{\rm onner}^{\rm model}$ &
$10.0\pm 2.1$ &
$9.2^{+2/6}_{-1.5}$ &
$8.8^{+1.7}_{-1.3}$ &
$9.2^{+1.7}_{-1.4}$ \\
$\Phi_{\rm resolved}^{\rm model}$ ~~[$10^{-11}$\;(cm$^2$s)$^{-1}$] & 
$	3.3^{+0.7}_{-0.5} $ 
&
$	4.2	^{+	1.4	}_{	-0.7	}$
&
$	5.1	^{+	8.7	}_{	-2.0	}$
&
$	5.2	^{+	4.7	}_{	-0.7	}$ \\
$\Phi_{\rm unresolved}^{\rm model}$ ~~[$10^{-11}$\;(cm$^2$s)$^{-1}$] & 
$	0.18	^{+	0.19	}_{	-0.08	}$
&
$	0.16	^{+	0.09	}_{	-0.08	}$
&
$	0.28	^{+	0.22	}_{	-0.11	}$
&
$	0.54	^{+	0.64	}_{	-0.23	}$
\\
$f_u = \Phi_{\rm unresolved}^{\rm model}/\Phi_{\rm all}^{\rm model}$ & 
$	0.055	^{+	0.075	}_{	-0.029	}$
&
$	0.039	^{+	0.034	}_{	-0.023	}$
&
$	0.054	^{+	0.100	}_{	-0.042	}$
&
$	0.100	^{+	0.160	}_{	-0.073	}$
\\
\hline
\end{tabular}
\end{table}

%%%%%%%%%%%%%%%%%%%%%%%%%%%%%%%%%%%%%%%%%%%%%%%%%%%%%%%%%%%%%%%%%%%%%%%%%%%%
%% Table. LHAASO-KM2A (25-100 TeV)
%%%%%%%%%%%%%%%%%%%%%%%%%%%%%%%%%%%%%%%%%%%%%%%%%%%%%%%%%%%%%%%%%%%%%%%%%%%%

\begin{table}
\caption{\footnotesize
Estimate of gamma--ray sources properties obtained
from the observations of the LHAASO--KM2A telescope  \cite{LHAASO:2023rpg}
in the energy range [25,100]~TeV.
The structure of this table is the same as for
table~\ref{tab:fit_hawc} and~\ref{tab:fit_wcda} with the top part 
listing experimental results, and the bottom part
showing theoretical estimates obtained
using the same models.
Experimental results and theoretical estimates
are given for the two sky regions (Inner--Galaxy and Outer--Galaxy region)
where LHAASO has published measurements of the diffuse flux.
Four rows compare the model estimates (obtained with
two different methods, see main text) of the unresolved source flux in the two (Inner--Galaxy and Outer--Galaxy) regions
with the measurements of the diffuse gamma--ray flux.
\label{tab:fit_km2a}}

\vspace{0.35cm}
\renewcommand{\arraystretch}{1.4}
\begin{tabular}{ | l || c | c | c | c |}
 \hline
\multicolumn{5}{|l|}
{LHAASO--KM2A Telescope. Energy Interval [25,100]~TeV} \\
\multicolumn{5}{|l|}
{All Sky: $N_{\rm sources} = 75$ ~~~~Region A: $N_{A} = 58$
 ~~~~Region B: $N_{B} = 17$. } \\
\hline
\multicolumn{5}{|l|} {LHAASO Inner--Galaxy
 [($|b|< 5^\circ$), ($15^\circ \le \ell \le 125^\circ$) } \\
\multicolumn{5}{|l|} {$N_{\rm sources}^{\rm data} = 52$} \\
\multicolumn{5}{|l|} {$\Phi_{\rm resolved}^{\rm data} = 11.9 \times 10^{-13}$ ~[cm$^{-2}$s$^{-1}$].} \\
\multicolumn{5}{|l|} {$\Phi_{\rm diffuse}^{\rm data} = (3.12^{+0.42}_{-0.39}) \times 10^{-13}$ ~[cm$^{-2}$s$^{-1}$]
 ~~~ (masking sources)} \\ 
\multicolumn{5}{|l|} {$\Phi_{\rm diffuse}^{\rm data}/\Phi_{\rm resolved}^{\rm data} = 0.26\pm 0.03$} \\
\hline
\multicolumn{5}{|l|} {LHAASO Outer--Galaxy
 [($|b|< 5^\circ$), ($125^\circ \le \ell \le 235^\circ$)] } \\
\multicolumn{5}{|l|} {$N_{\rm sources}^{\rm resolved} = 10$} \\
\multicolumn{5}{|l|} {$\Phi^{\rm resolved}_{\rm sources} = 1.53 \times 10^{-13}$ ~[cm$^{-2}$s$^{-1}$]} \\
\multicolumn{5}{|l|} {$\Phi_{\rm diffuse}^{\rm data} = (1.38^{+0.29}_{-0.26}) \times 10^{-13}$ ~[cm$^{-2}$s$^{-1}$]
~~~ (masking sources)} \\
\multicolumn{5}{|l|} {$\Phi_{\rm diffuse}^{\rm data}/\Phi^{\rm resolved}_{\rm sources} = 0.90^{+0.19}_{-0.17}$} \\

\hline
\hline
\hline

\multicolumn{5}{|l|} {Modeling of gamma--ray sources} \\
\hline
~~ & Model~1 & Model~2 & Model~3 & Model 4 \\
\hline
Luminosity distribution & Identical sources & Identical sources &
P.L. ($\gamma = 1.25$) & P.L ($\gamma =1.25$) \\
Space distribution & Smooth (PWN) & Spirals & Smooth (PWN) & Smooth (PWN) \\
Source size & $R = 0$ & $R = 0$ & $R = 0$ & $R= 20$~pc \\ 
\hline
\multicolumn{5}{|l|} {Global properties of Galactic gamma--ray sources} \\
\hline
Horizon [kpc] & 
$	7.9	^{+	2.4	}_{	-2.5	}$
&
$	8.1	^{+	1.9	}_{	-3.2	}$
&
$	18.8	^{+	18.	}_{	-9.	}$
&
$	14.2	^{+	11.	}_{	-7.	}$
\\
$L_0$ or $L_*$ [$10^{32}$~erg/s] & 
$	6.9	^{+	4.9	}_{	-3.7	}$
&
$	7.2	^{+	3.7	}_{	-4.6	}$
&
$	39^{+80}_{	-29}$
&
$	22.	^{+	49.	}_{	-16.	}$
\\$L_{\rm tot}$ [$10^{35}$~erg/s] & 
$	5.5	^{+	0.75	}_{	-0.60	}$
&
$	4.5	^{+	0.56	}_{	-1.3	}$
&
$	7.2	^{+	4.8	}_{	-1.9	}$
&
$	8.7	^{+	2.3	}_{-0.5	}$
\\
\hline
\hline
\multicolumn{5}{|l|} {Modeling LHAASO--KM2A  Inner--Galaxy region } \\
\hline
$N_{\rm Inner}^{\rm model}$ &
$49^{+5}_{-9}$ & $48^{+5}_{-9}$ & $48^{+5}_{-9}$ &$48^{+5}_{-9}$ 
\\
$\Phi_{\rm resolved}^{\rm model}$~~~[$10^{-13}$\;(cm$^2$s)$^{-1}$] & 
$	4.2	^{+	1.2	}_{	-1.2	}$
&
$	4.8	^{+	1.1	}_{	-2.1	}$
&
$	6.2	^{+	5.9	}_{	-2.7	}$ 
&
$	6.2	^{+	3.6	}_{	-1.9	}$ \\
$\Phi_{\rm unresolved}^{\rm model}$[$10^{-13}$\;(cm$^2$s)$^{-1}$] & 
$	1.9	^{+	0.6	}_{	-0.4	}$
&
$	1.5	^{+	0.4	}_{	-0.4	}$
&
$	1.9	^{+	0.6	}_{	-0.6	}$
&
$	3.3	^{+	2.4	}_{	-1.1	}$
\\
$f_u = \Phi_{\rm unresoled}^{\rm model}/\Phi_{\rm all}^{\rm model}$ & 
$	0.31	^{+	0.14	}_{	-0.09	}$
&
$	0.24	^{+0.16	}_{-0.08	}$
&
$	0.23	^{+	0.19	}_{-0.13	}$
&
$	0.35	^{+	0.22	}_{	-0.16	}$ \\
\hline
$\Phi_{\rm unresolved}^{\rm model}/\Phi_{\rm diffuse}^{\rm data}$ & 
$0.89^{+0.09}_{-0.30}$ &
$0.70^{+0.08}_{-0.26}$ &
$0.87^{+0.12}_{-0.32}$ &
$1.6^{+0.7}_{-0.6}$ 
\\
$[\Phi_{\rm sources}^{\rm data} \, f_u/(1-f_u)] /\Phi_{\rm diffuse}^{\rm data}$ ~~&
$2.5^{+1.4}_{-1.2}$ &
$1.7^{+1.4}_{-0.9}$ &
$1.7^{+1.7}_{-1.1}$ &
$3.0^{+3.2}_{-1.9}$ \\
\hline
\hline
\multicolumn{5}{|l|} {Modeling LHAASO--KM2A  Outer--Galaxy region. } \\
\hline
$N_{\rm Outer}^{\rm model}$ &
$13\pm 2$ & $12\pm 2$ & $11^{+2}_{-1}$ & $12^{+2}_{-1}$ 
\\
$\Phi_{\rm resolved}^{\rm model}$~~[$10^{-13}$\;(cm$^2$s)$^{-1}$] & 
$	2.1	\pm 0.4 $ 
&
$	2.9	^{+	0.4	}_{	-1.0	}$
&
$	2.8	^{+	2.1	}_{	-1.0	}$
&
$	3.0	^{+	1.2	}_{	-0.5	}$ \\
$\Phi_{\rm unresolved}^{\rm model}$ [$10^{-13}$\;(cm$^2$s)$^{-1}$] & 
$	0.18	^{+	0.15	}_{	-0.07	}$
&
$	0.14	^{+	0.15	}_{	-0.04	}$
&
$	0.26	^{+	0.17	}_{	-0.09	}$
&
$	0.64	^{+	0.7	}_{	-0.3	}$
\\
$f_u = \Phi_{\rm unresolved}^{\rm model}/\Phi_{\rm all}^{\rm model}$ & 
$	0.077	^{+	0.081	}_{	-0.035	}$
&
$	0.047	^{+	0.086	}_{	-0.016	}$
&
$	0.088	^{+	0.110	}_{	-0.052	}$
&
$	0.18	^{+	0.17	}_{	-0.095	}$
\\
\hline
$\Phi_{\rm unresolved}^{\rm model}/\Phi_{\rm diffuse}^{\rm data}$ & 
$0.13^{+0.07}_{-0.06}$ &
$0.10^{+0.07}_{-0.04}$ &
$0.19^{+0.07}_{-0.09}$ &
$0.47^{+0.35}_{-0.24}$ 
\\
$[\Phi_{\rm sources}^{\rm data} \, f_u/(1-f_u)] /\Phi_{\rm diffuse}^{\rm data}$ ~~&
$0.09^{+0.08}_{-0.05}$ &
$0.05^{+0.09}_{-0.03}$ &
$0.11^{+0.12}_{-0.07}$ &
$0.24^{+0.27}_{-0.16}$ \\
\hline
\end{tabular}
\end{table}

%%%%%%%%%%%%%%%%%%%%%%%%%%%%%%%%%%%%%%%%%%%%%%%%%%%%%%%%%%%%%%%%%%%%%%%%%%%%
%% Table. LHAASO-KM2A (100-1000 TeV)
%%%%%%%%%%%%%%%%%%%%%%%%%%%%%%%%%%%%%%%%%%%%%%%%%%%%%%%%%%%%%%%%%%%%%%%%%%%%

\begin{table}
\caption{\footnotesize
Estimate of gamma--ray sources properties obtained
from the observations of the LHAASO--KM2A telescope \cite{LHAASO:2023rpg}
in the energy range [100,1000]~TeV.
The structure of this table is the same as for
table~\ref{tab:fit_km2a} with the top part 
listing experimental results, and the bottom part
showing theoretical estimates obtained using the same models.
Four rows compare the model estimates (obtained with
two different methods, see main text) of the unresolved source
flux in the two (Inner--Galaxy and Outer--Galaxy) sky regions
with the measurements of the diffuse gamma--ray flux.
\label{tab:fit_km2a-hig}}

\vspace{0.35cm}
\renewcommand{\arraystretch}{1.4}
\begin{tabular}{ | l || c | c | c | c |}
 \hline
\multicolumn{5}{|l|}
{LHAASO--KM2A Telescope. Energy Interval [100--1000]~TeV} \\
\multicolumn{5}{|l|}
{All Sky: $N_{\rm sources} = 44$ ~~~~Region A: $N_{A} = 36$ ~~~
 ~~~~Region B: $N_{B} = 8$. } \\
\hline
\multicolumn{5}{|l|} {LHAASO Inner--Galaxy
 [($|b|< 5^\circ$), ($15^\circ \le \ell \le 125^\circ$) } \\
\multicolumn{5}{|l|} {$N_{\rm sources}^{\rm data} = 30$} \\
\multicolumn{5}{|l|} {$\Phi^{\rm resolved}_{\rm sources} = 3.70 \times 10^{-14}$ ~[cm$^{-2}$s$^{-1}$].} \\
\multicolumn{5}{|l|} {$\Phi_{\rm diffuse}^{\rm data} = (2.09^{+0.43}_{-0.37}) \times 10^{-14}$ ~[cm$^{-2}$s$^{-1}$]
 ~~~ (masking sources)} \\ 
\multicolumn{5}{|l|} {$\Phi_{\rm diffuse}^{\rm data}/\Phi^{\rm resolved}_{\rm sources} = 0.56^{+0.12}_{-0.10}$} \\
\hline
\multicolumn{5}{|l|} {LHAASO Outer--Galaxy
 [($|b|< 5^\circ$), ($125^\circ \le \ell \le 235^\circ$)] } \\
\multicolumn{5}{|l|} {$N_{\rm sources}^{\rm resolved} = 3$} \\
\multicolumn{5}{|l|} {$\Phi^{\rm resolved}_{\rm sources} = 0.243 \times 10^{-14}$ ~[cm$^{-2}$s$^{-1}$]} \\
\multicolumn{5}{|l|} {$\Phi_{\rm diffuse}^{\rm data} = (0.92^{+0.32}_{-0.25}) \times 10^{-14}$ ~[cm$^{-2}$s$^{-1}$]
~~~ (masking sources)} \\
\multicolumn{5}{|l|} {$\Phi_{\rm diffuse}^{\rm data}/\Phi^{\rm resolved}_{\rm sources} = 3.8^{+1.2}_{-1.0}$} \\

\hline
\hline
\hline

\multicolumn{5}{|l|} {Modeling of gamma--ray sources} \\
\hline
~~ & Model~1 & Model~2 & Model~3 & Model 4 \\
\hline
Luminosity distribution & Identical sources & Identical sources &
P.L. ($\gamma = 1.25$) & P.L ($\gamma =1.25$) \\
Space distribution & Smooth (PWN) & Spirals & Smooth (PWN) & Smooth (PWN) \\
Source size & $R = 0$ & $R = 0$ & $R = 0$ & $R= 20$~pc \\ 
\hline
\multicolumn{5}{|l|} {Global properties of Galactic gamma--ray sources} \\
\hline
Horizon [kpc] & 
$	10.7	^{+	5.2	}_{	-4.6	}$
&
$	10.4	^{+	7.5	}_{	-3.8	}$
&
$	41.5	^{+	90.	}_{	-30-.	}$
&
$	28.1	^{+	80.	}_{	-19.	}$
\\
$L_0$ or $L_*$ [$10^{32}$~erg/s] & 
$	2.6	^{+	3.1	}_{	-1.7	}$
&
$	2.4	^{+	4.7	}_{	-1.4	}$
&
$	38^{+80}_{	-29}$
&
$	18.	^{+	80.	}_{	-16.	}$
\\
$L_{\rm tot}$ [$10^{34}$~erg/s] & 
$	7.6	^{+	2.9	}_{	-1.6	}$
&
$	6.3	^{+	5.2	}_{	-1.5	}$
&
$	16.0	^{+	15	}_{	-9.1	}$
&
$	13.9	^{+	12	}_{	-3.6	}$
\\
\hline
\hline
\multicolumn{5}{|l|} {Modeling LHAASO--KM2A ($E > 100$~TeV)  Inner--Galaxy region } \\
\hline
$N_{\rm Inner}^{\rm model}$ &
$32^{+3}_{-7}$ & $31^{+3}_{-6}$ & $31^{+2}_{-7}$ &$32^{+2}_{-7}$ 
\\
$\Phi_{\rm resolved}^{\rm model}$~~~[$10^{-14}$\;(cm$^2$s)$^{-1}$] & 
$	1.5	^{+	0.9	}_{	-0.6	}$
&
$	1.7	^{+	1.9	}_{	-0.7	}$
&
$	3.6	^{+	7	}_{	-2.5	}$ 
&
$	2.9	^{+	9 }_{	-1.6	}$ \\
$\Phi_{\rm unresolved}^{\rm model}$[$10^{-14}$\;(cm$^2$s)$^{-1}$] & 
$	0.39	^{+	0.22	}_{	-0.13	}$
&
$	0.29	^{+	0.18	}_{	-0.13	}$
&
$	0.36	^{+	0.25	}_{	-0.17	}$
&
$	0.57	^{+	0.71	}_{	-0.39	}$
\\
$f_u = \Phi_{\rm unresoled}^{\rm model}/\Phi_{\rm all}^{\rm model}$ & 
$	0.20	^{+	0.20	}_{	-0.13	}$
&
$	0.15	^{+0.21	}_{-0.09}$
&
$	0.09	^{+	0.27	}_{-0.09	}$
&
$	0.16	^{+	0.33	}_{	-0.12	}$ \\
\hline
$\Phi_{\rm unresolved}^{\rm model}/\Phi_{\rm diffuse}^{\rm data}$ & 
$0.19^{+0.27}_{0.13}$ &
$0.14^{+0.21}_{-0.09}$ &
$0.17^{+0.27}_{-0.09}$ &
$0.27^{+0.47}_{-0.03}$ 
\\
$[\Phi_{\rm sources}^{\rm data} \, f_u/(1-f_u)] /\Phi_{\rm diffuse}^{\rm data}$ ~~&
$0.59^{+0.13}_{-0.13}$ &
$0.40^{+0.09}_{-0.09}$ &
$0.23^{+0.6}_{-0.2}$ &
$0.45^{+0.7}_{-0.4}$ \\
\hline
\hline
\multicolumn{5}{|l|} {Modeling LHAASO--KM2A ($E> 100$~TeV)  Outer--Galaxy region. } \\
\hline
$N_{\rm Outer}^{\rm model}$ &
$6.3^{+2.0}_{-1.4}$ &
$5.8^{+2.2}_{-1.2}$ &
$5.6^{+1.6}_{-1.1}$ &
$5.8^{+1.6}_{-1.2}$ 
\\
$\Phi_{\rm resolved}^{\rm model}$~~[$10^{-14}$\;(cm$^2$s)$^{-1}$] & 
$	0.70^{+0.29}_{-0.19}$ &
$	0.92	^{+	0.81	}_{	-0.19	}$
&
$	1.5	^{+	0.80	}_{	-0.25	}$
&
$	1.2	^{+	0.72	}_{	-0.53	}$ \\
$\Phi_{\rm unresolved}^{\rm model}$ [$10^{-14}$\;(cm$^2$s)$^{-1}$] & 
$	0.028	^{+	0.045	}_{	-0.015	}$
&
$	0.026	^{+	0.025	}_{	-0.018	}$
&
$	0.045	^{+	0.055	}_{	-0.022	}$
&
$	0.0.094	^{+	0.19	}_{	-0.067	}$
\\
$f_u = \Phi_{\rm unresolved}^{\rm model}/\Phi_{\rm all}^{\rm model}$ & 
$	0.038	^{+	0.088	}_{	-0.025	}$
&
$	0.028	^{+	0.043	}_{	-0.034	}$
&
$	0.029	^{+	0.12	}_{	-0.029	}$
&
$	0.071	^{+	0.22	}_{	-0.067	}$
\\
\hline
$\Phi_{\rm unresolved}^{\rm model}/\Phi_{\rm diffuse}^{\rm data}$ & 
$0.030^{+0.050}_{-0.014}$ &
$0.029^{+0.050}_{-0.01-}$ &
$0.048^{+0.070}_{-0.031}$ &
$0.100^{+0.180}_{-0.062}$ 
\\
$[\Phi_{\rm sources}^{\rm data} \, f_u/(1-f_u)] /\Phi_{\rm diffuse}^{\rm data}$ ~~&
$0.016^{+0.014}_{-0.011}$ &
$0.012^{+0.01-}_{-0.010}$ &
$0.012^{+0.33}_{-0.010}$ &
$0.31^{+0.068}_{-0.06}$ \\
\hline
\end{tabular}
\end{table}

\clearpage

%%%%%%%%%%%%%%%%%%%%%%%%%%%%%%%%%%%%%%%%%%%%%%%%%%%%%%%%%%%%%%%%%%%%%%%%%%%%
%% Figures
%%%%%%%%%%%%%%%%%%%%%%%%%%%%%%%%%%%%%%%%%%%%%%%%%%%%%%%%%%%%%%%%%%%%%%%%%%%%

%%%%%%%%%%%%%%%%%%%%%%%%%%%%%%%%%%%%%%%%%%%%%%%%%%%%%%%%%%%%%%%%%%

%%%%%%%%%%%%%%%%%%%%%%%%%%%%%%%%%%%%%%%%%%%%%%%%%%%%%%%%%%%%%%%%%
% Measurements of the diffuse flux
%---------------------------------------------------------------

\clearpage

%%%%%%%%%%%%%%%%%%%%%%%%%%%%%%%%%%%%%%%%%%%%%%%%%%%%%%%%%%%%%%%%%%%%%%%%%%%%
% Geometry of space distributions 
%---------------------------------------------------------------------------

%%%%%%%%%%%%%%%%%%%%%%%%%%%%%%%%%%%%%%%%%%%%%%%%%%%%%%%%%%%%%%%%%%%%%%%%%%%%
% Luminosity distribution 
%--------------------------------------------------------------------------

\clearpage

\clearpage

%%%%%%%%%%%%%%%%%%%%%%%%%%%%%%%%%%%%%%%%%%%%%%%%%%%%%%%%%%%%%%%%%
%%% HAWC sources + diffuse flux
%----------------------------------------------------------------

%%%%%%%%%%%%%%%%%%%%%%%%%%%%%%%%%%%%%%%%%%%%%%%%%%%%%%%%%%%%%%%%%
%%% LHAASO sources + diffuse flux
%----------------------------------------------------------------

\clearpage

%%%%%%%%%%%%%%%%%%%%%%%%%%%%%%%%%%%%%%%%%%%%%%%%%%%%%%%%%%%%%%%%%
% IceCube discussion 
%---------------------------------------------------------------

\clearpage

\clearpage

%%%%%%%%%%%%%%%%%%%%%%%%%%%%%%%%%%%%%%%%%%%%%%%%%%%%%%%%%%%%%%%%%%%%%%

%\bibliographystyle{plain}
%\bibliography{abib}
 
%%%%%%%%%%%%%%%%%%%%%%%%%%%%%%%%%%%%%%%%%%%%%%%%%%%%%%%%%%%%%%%%%%%%%%

\end{document}